\documentstyle[12pt,aasms]{article}

\lefthead{??}
\righthead{}

\def\lta{\lesssim}
\def\gta{\gtrsim}
\def\Icorr{I_{improve}}

\def\lta{{\>\rlap{\raise2pt\hbox{$<$}}\lower3pt\hbox{$\sim$}\>}}
\def\gta{{\>\rlap{\raise2pt\hbox{$>$}}\lower3pt\hbox{$\sim$}\>}}

\def\ltsima{$\; \buildrel < \over \sim \;$}
\def\simlt{\lower.5ex\hbox{\ltsima}}

\begin{document}

\title{Ellipticals with Kinematically-Distinct
 Cores: $V-I$ Color Images with {\tt WFPC2}$^1$}

\author{C. Marcella Carollo$^2$}

\affil{The Johns Hopkins University, Bloomberg Center, 21218 MD, Baltimore;
marci@pha.jhu.edu}

\author{Marijn Franx}

\affil{Kapteyn Laboratory, University of Groningen, P.O. Box 800, 
9700 AV Groningen, \\The Netherlands; franx@astro.rug.nl}

\author{Garth D. Illingworth and Duncan A. Forbes}

\affil{Lick Observatory, University of California, Santa Cruz, CA 95064;\\
gdi@ucolick.org, forbes@ucolick.org}

\vspace*{3truecm}

$^1$ Based on observations
with the NASA/ESA Hubble Space Telescope, obtained at the Space Telescope
Science Institute, which is operated by Association of Universities for
Research in Astronomy, Inc. (AURA), under NASA contract NAS5-26555.

$^2$ Hubble Fellow

\begin{abstract}

We have analysed HST {\tt WFPC2} F555W and F814W (i.e., $V$ and $I$)
images for fifteen elliptical galaxies with kinematically-distinct cores. For
each of them we have derived  surface brightness and isophotal
parameter profiles in the two bands, color maps, and radial
profiles in $V-I$.  
Most galaxies show patchy dust absorption close to their nuclei.
However, there are generally no indications of
 homogeneous, diffuse dust components close to the nuclei.
The  nuclear colors in the unobscured regions are  most likely
representative of the central stellar populations.

We have detected photometric evidence for faint stellar disks, on
scales of a few tens to a few arcseconds, in seven galaxies, namely
NGC 1427, 1439, 1700, 4365, 4406, 4494 and 5322. In NGC 1700, the
isophotes are slightly boxy at the scale of the counter-rotating
component, and disky at larger radii. We find no difference in $V-I$
color greater than 0.02 mag between these disks and the surrounding
galactic regions.  Hence the stellar populations in the kinematically
distinct cores are not strongly deviant from the population of the
main body.  Specifically, there is no evidence for a dominating
population of blue, very metal weak stars, as predicted by some of the
formation scenarios.  This argues against models in which small
galaxies fall in and survive in the nuclei, unless super massive black
holes are present. These would in fact disrupt the accreted small
systems.  

For one galaxy, NGC 4365, the innermost region is
bluer than the surrounding regions.
This area extends to  $\sim 15$pc, and contains a 
luminosity of  $\sim 2.5 \times 10^6$ L$_\odot$. 
If interpreted as a stellar
population effect, an age difference of $\sim$ 3-4 Gyrs, or an
$[Fe/H]$ variation of about 0.2 dex, is derived.

The nuclear intensity profiles show a large variety: some galaxies
have steep cusp profiles, others have shallow cusps and a ``break
radius''.  The nuclear cusps of galaxies with kinematically-distinct
cores follow the same trends as the nuclei of normal galaxies.

We have not been able to identify a unique, qualifying feature in the
{\tt WFPC2} images which distinguish the galaxies with kinematically distinct
cores from the kinematically normal cores.
It is possible that statistical differences exist: possibly,  the
kinematically distinct cores have a higher fraction of nuclear disks.
The similarity of both types of cores puts strong constraints on the
formation scenarios. Simulations of galaxy mergers, with the inclusion
of star formation and nuclear black holes, are needed to resolve
the question how these structures may have formed.
Spectra with high spatial resolution are needed to study the nuclear
structure of the distinct component in detail.

\end{abstract}

\section{Introduction}

The study of nearby (kinematically-normal) elliptical galaxies
performed with the {\tt WF/PC} camera aboard the {\it Hubble Space
Telescope} (HST) has revealed the presence of central disks, dust
clouds, and nuclear components on scale lengths of a few tens of
parsecs (e.g., Jaffe et al. 1994; van den Bosch et al. 1994; Lauer et
al. 1995, hereafter L95; van Dokkum \& Franx 1995).  In all galaxies,
the surface brightness profile $I(r)$ for $r \rightarrow 0.1''$ can be
approximated as a power--law $r^{-\gamma}$, with $0\lta \gamma \lta
1$.  Galaxies with steep cusps are almost always small, fast--rotating
objects, while in contrast massive, pressure--supported ones have
shallow cusps, typically with $\gamma \lta 0.5$ (e.g., L95; Kormendy
et al. 1995, hereafter K95).  It is not clear whether this reflects a
real dichotomy in the physical properties and formation processes, as
suggested by L95 and K95. In any case, nuclear properties on scales of
a few tens of parsecs are correlated with properties observed on
global scales, and it is legitimate to ask whether this is the outcome
of selection effects acting at formation, or whether it has resulted
from subsequent galactic evolution.

Galaxies with kinematically-distinct cores are very interesting
physical laboratories to test ideas of galaxy formation and evolution.
Their distinct angular momenta suggest that these cores may be the
relics of interactions or mergers, and thus may provide a powerful
diagnostic to estimate the relevance of interactions in shaping
galactic properties.  Kormendy (1984) argued that the core of NGC 5813
is dominated by the remnant of a small galaxy that fell into a big
elliptical. This scenario predicts core/global differences that should
be detectable not only in the kinematics, but also in the stellar
populations and overall structure.  Another possible scenario for
kinematically-distinct cores implies gas-rich mergers of two galaxies,
where the gas is converted into stars in the central regions, and
subsequent star formation produces a disk which retains the orbital
signature of the original gas.  Still another possibility is that
formation of kinematically-distinct cores is a rather normal aspect of
a hierarchical formation scenario for elliptical galaxies, where
galaxies are built up from pre-existing clumps (possibly different
from current galaxies). The last two scenarios do not necessarily
predict any measurable difference in the stellar population of the
distinct core.

Forbes, Franx \& Illingworth (1995, hereafter F95) investigated
 HST {\tt WF/PC} F555W images for eight 
early-type galaxies with kinematically-distinct cores.
These authors found clear indications of small-scale dust in
six out of the eight galaxies, and argued that the dust might
be directly linked to the existence of nuclear radio activity in
these galaxies. They also showed that the nuclear surface brightness
profiles in this class of galaxies are similar to those 
in kinematically-normal galaxies. On this basis, they argued against the
plausibility of the small galaxy accretion scenario, since this low-mass
object should dominate the light in the core region and produce differences
with respect to normal galaxies (that are not observed). 
F95 searched for the nuclear disks expected in the dissipative 
core formation scenario (and suggested by spectroscopic analysis 
of several distinct cores, see e.g., Franx \& Illingworth 1988). The search
lead to one detection in NGC 4365
(previously reported by Forbes 1994), and to a couple of other candidates.
In the other galaxies, no conclusions could be drawn due to 
the strong dust absorption.

Additional information is clearly needed to resolve the enigma of the
formation of kinematically-distinct cores.  Multi-color,
high-resolution imaging can provide a valuable key for understanding
the origin of these peculiar stellar systems.  In particular, within
the limits imposed by their degeneracy between age and metallicity,
{\it (i)} color gradients between the core region and the rest of the
galaxy could be used to estimate metallicity/age differences in the
stellar populations; {\it (ii)} color information could help to detect
nuclear disks, and indicate whether such disks were formed at
the same time as the majority of stars (or more recently, as a bluer
color might indicate); {\it (iii)} colors of the central cusps could
provide insight into their formation, discriminating between accretion
of low-mass stellar systems (possibly leading to bluer cusps) and
stars formed in situ (possibly leading to red cusps).

Observationally, colors and color gradients at subarcsecond scales are
practically unexplored for any kind of galaxy.  The only other example
to date is that of Crane et al. (1993), who measured gradients for 12
galaxies using pre-refurbishment FOC data, and concluded that color
variations are either very small or absent in galactic
nuclei. However, these measurements are limited by low S/N and the
uncertain {\tt pre-COSTAR} PSF.

In this paper we extend the investigation started by F95 by studing
{\tt WFPC2} images in two bands, namely F555W and F814W (i.e., similar
to Johnson $V$ and Cousins $I$, respectively), for 15 galaxies with
kinematically-distinct cores.  The new sample contains the eight
galaxies of F95, plus seven additional galaxies that also show
peculiar core kinematics.  Eleven of the galaxies are luminous objects
with absolute $V$ mag $M_V\le$-20 + 5 log $h$ ($h=H_o/100$ km s$^{-1}$
Mpc$^{-1}$ throughout this paper).  The sample, the observations and
the basic photometric analysis are described in Section 2.  The $V$
and $I$ surface brightness profiles and the parameters defining the
core morphology are presented in Section 3, together with the $V-I$
color profiles, the color maps, and the color gradients. In Section 4
we present the adopted parametrization of the surface brightness
profiles, and derive the global estimator of the nuclear
cusp-steepness which we will use in our discussion.  In Section 5 we
discuss the main results of this study, and summarize them in (the
concluding) Section 6.  In four appendices, we describe {\it (i)} the
procedure used to minimize the effects of nuclear patchy dust on the
derivation of the surface brightness profiles, {\it (ii)} the
comparison of our {\tt WFPC2} measurements with {\tt WF/PC} profiles
already published for some galaxies of our sample; {\it (iii)} the
deprojections of the surface density profiles, performed in order to
obtain an estimate of the physical luminosity density and of its
nuclear radial variation; and {\it (iv)} the derivation of the nuclear
properties of the L95 galaxies, performed so as to compare
kinematically-peculiar with normal galaxies. The properties of the
globular cluster systems in our sample are derived and discussed by
Forbes et al. (1996).

\section{Observations and Measurements}

\subsection{The Sample}

The 15 galaxies constituting the sample are listed in Table 1,
together with some global parameters. The galaxies cover a large range
in optical, infrared, X-ray and radio luminosities, and in
metallicities (as characterized by central Mg$_2$ index values).
Seven of them (NGC 1700, 3608, 4278, 4589, 5322, 5982 and 7626) belong
to the sample of Schweizer et al. (1990), and cover a large range in
the ``fine-structure'' $\Sigma$ parameter (from a value of 3.7 for NGC
1700, to zero for NGC 3608 and NGC 4589).  The environment of the
galaxies varies considerably, with some galaxies closely paired to a
companion in a rather loose group (e.g., IC 1459 and NGC 7626), and
others embedded in rich clusters (e.g., the Virgo galaxy NGC 4406).

\subsection{Data Reduction}

For each galaxy we acquired two F555W exposures of 500 seconds each,
and two F814W exposures of 230 seconds each.
The observations were taken with the WFPC2 camera centered on PC1,
with its scale of 0.046$''$ px$^{-1}$,
and a field of about $36''\times36''$.

Initial data reduction was performed using the standard STScI pipeline
software.  For each filter, the two available images were compared to
each other, and found to have the same central pixel value within
Poisson statistics. Their relative alignment was checked before
combining them, and found to be better than about 0.2 px.  The two
images were averaged and cosmic-ray cleaned using the cosmic ray
rejection option of the STSDAS task ``combine'' , in combination with
a noise model appropriate for the PC1 (all the images were taken with
gain 15, except those for NGC 1439, 4406, 4552 and 5322, which were
taken with gain 7).  Hot and cold pixels were partly removed as well
by the combining task; an interpolation over the remaining defects was
subsequently performed.  The cosmic-ray and bad-pixel removal was
checked by inspecting the residual difference between the cleaned
image and each of the two original frames.  Except for NGC 1700, the
observations were taken after 1994 April 24.  We did not correct for
charge transfer efficiency (CTE), since this is relevant only for
point sources in low background (Stiavelli 1996),
and also given the uncertainties associated with the available
algorithm (Holtzman et al. 1995, hereafter H95).  In the case of NGC
1700, we forced the isophotes to remain concentric over the whole
radial range, and checked {\it a posteriori} for any bias in the
measurements.

Sky subtraction was carried out for some of the galaxies, although the 
sky level was very small (smaller than 1DN).  The sky signal was determined 
from the WFC images, in areas furthermost from the nucleus.  No sky was
subtracted from NGC 4406 and NGC 4552, since these galaxies obviously 
filled the field of view. The uncertainties introduced by not making this 
correction, or any contamination by the galaxy of the ``sky'' regions,
were estimated to be negligible.  This was verified by comparison
with ground-based photometry in the overlap region.  Good agreement 
in the radial variation of the surface brightness was found between 
our outer surface brightness profiles and the ground-based photometry 
published by Sparks et al. (1991), Goudfrooij et al. (1994a), and 
M{\o}ller et al. (1995).

The photometric zero-point was derived and converted to the
Johnson-Cousins system following the iterative procedure described in
H95.  For all the galaxies, we first set
\begin{eqnarray}
mag_j^o &  = & -2.5  Log(counts/sec)+Z_{WFPC} \\
   & & +2.5Log(GR)+5 Log(0.046)+0.1\nonumber
\end{eqnarray}
with the subscript $j$ indicating either the $V$ or the $I$ filter,
 $Z_{WFPC}$ the zero point of the filter (equal to 21.724 for F555W,
and to 20.840 for F814W; see Table 7 of H95), and GR the gain 
ratio (set to 1 for gain 15, and to 1.987 for gain 7). 
The subsequent term accounts for the pixel size. The constant shift of 0.1 
mag was added following H95 in order to correct for infinite aperture. 
Then we iterated the calibration until convergence by inserting
the newly derived color term in Eq. (8) of H95, i.e.:
\begin{equation}
mag_j= mag_j^o +c_1 (V-I)+c_2 (V-I)^2
\end{equation}
with $c_1=-0.052$ and $c_2=0.027$ for the $V$ filter, and $c_1=-0.063$
and $c_2=0.025$ for the $I$ filter (Table 7 of H95).  The final
surface brightness profiles were corrected for Galactic extinction
(see Table 1). K-corrections were also applied (Whitford 1971), though
they were always smaller than 0.03 mag.

Comparison with the ground-based photometry of Goudfrooij et
al. (1994a) and M{\o}ller et al. (1995) showed small zeropoint offsets
in some cases.  These were generally smaller than $\sim 0.1$ mag.
There was no detectable systematic difference in the $V-I$ colors
between our HST measurements and the ground-based results
($\Delta(V-I) \lta 0.05$ mag).

\subsection{Ellipse Fitting Procedure}

We modeled each galaxy by means of an iterative procedure which fits
isophotes to both $V$ and $I$ final images.  The software package used
was {\it galphot} (J{\o}rgensen et al. 1992).  For all the galaxies,
fits to derive the isophotal morphological parameters were performed
on the $V$ and $I$ frames, after masking out stars and other
point-like sources. Most of the galaxies of our sample show extensive
dust patches close to their centers. In order to minimize their
effects on the determination of the  surface brightness
profiles in the heavily obscured nuclei, we adopted a modified version
of the procedure described by F95. These authors had
only one filter ($V$) available.  They determined a maximum radius of
influence for the dust (i.e., the ``dust radius'') and fixed the isophotal
properties (ellipticity and position angle) to a constant value inside
this radius. The values adopted were chosen equal to the one just
beyond the dust radius. The patchy-dust areas were then masked out,
and the surface brightness on each isophote (of fixed ellipticity and
position angle) inside the dust radius was finally computed. 

The availability of two filters allowed us to improve on the above
procedure.  We used a simple ``screen-model'' to obtain a rough
estimate of the dust absorption from the patchy variations of color
over the image (see Appendix A).  This model is very simplistic, and
is not more than a first order approximation to the complex effects of
dust.  None the less, visual inspection of the resulting
``dust-improved'' $I_{improve}$ images showed that the effects of
patchy dust were indeed decreased.  We used these images to obtain
``dust-improved'' values for the isophotal parameters, the ellipticity
and the position angle.  We tested whether these parameters were
sensitive to details of the model, such as the location of the screen.
We found no significant variations.

The values of ellipticity and position angle obtained from the fits to
the dust-improved $I_{improve}$ frames were then kept fixed in a
successive iteration of the ellipse fitting to the raw $I$ and $V$
images, in which the heavy dust patches had been previously masked
out. The so-obtained surface brightness profiles, which we will refer
to as to the {\it final dust-improved profiles}, were used in the
remainder of our analysis. Obviously, this procedure was not applied
to those galaxies devoid of heavy dust patches in their nuclei. For
these objects, the straight application of the isophotal fitting
package to the raw $I$ and $V$ images (masked for stars and other
spurious sources) already provided excellent fits, i.e. the 'final'
surface brightness profiles. More details on the procedure adopted for
deriving the surface brightness profiles in the heavily obscured
nuclei are given in Appendix A.

\subsection{Maps of $V-I$ Color}

The HST Point-Spread-Function (PSF) depends on position on the CCD.
Since no pointlike sources with adequate S/N located near to the
nuclei were available for any of the objects, we computed the
appropriate $V$ and $I$ PSF for each galaxy by running Tinytim (Krist
1992). This is the optimal procedure in such cases, given the fact
that archival stars can not be used due to the presence of focus
drifts (occurring on timescales of a few months), breathing (occurring
on timescales of a few minutes), and jitter (varying from exposure to
exposure).  Extensive use has been made of the Tinytim PSFs, and while
some concerns may remain, they are second order effects, and the use
of a first order correction is certainly better than none at all.

In order to derive the $(V-I)$ color maps and color profiles, the $V$
and $I$ final frames were matched to the same PSF by convolving each
band with the PSF of the other band (i.e., $V-I \equiv -2.5 log \Big(
{\frac {F^V_{conv,obs}}{F^I_{conv,obs}}} \Big)$, where the subscript
$conv$ stands for convolved images, and $F_{obs}$ are the counts in
the image frames).  The relative centering of the $V$ and $I$
convolved images was checked before generating the maps; in the case
of NGC 1427, we aligned the $V$ image to the $I$ image after rebinning
to a quarter of the pixel size. The PSF-matched color profiles were
derived by subtraction of the $V$ and $I$ surface brightness profiles
obtained by applying the same ellipse fitting procedure described in
Section 2.3 to the $V$ and $I$ PSF-convolved images.  Separate
measurements of $F^V$ and $F^I$ within apertures of $0.5''$, $1''$,
$2''$, and $5''$ of radius were made in order to obtain the $V-I$
color within those apertures.

The current data set does not allow us to derive accurate estimates
for dust extinction.  However, we used the method of Goudfrooij et
al. (1994b) to derive a first order estimate. We computed:

\begin{equation}
\label{adef}
A_\lambda = -2.5 log \Big( \frac {F^\lambda_{conv,obs}}{F^\lambda_{conv,mod}}  \Big),
\end{equation}

\noindent
where $\lambda\equiv I$ or $V$, and $F_{mod}$ is the number of counts
in the isophotal fit models of the PSF-convolved frames.  The mean
extinctions $A_{V,apert}$ and $A_{I,apert}$ inside an aperture of
$5''$ were computed after masking the pixels with $A_V \ge 0.03$ on
the $A_V$ and $A_I$ frames. The corresponding values of color excess
$E(V-I)_{apert} \equiv A_{V,apert}-A_{I,apert}$ were derived.  Typical
values are $E(V-I)_{apert} \simeq 0.02$ and $A_{V,apert} \simeq 0.05$,
with internal errors smaller than $\approx$ 0.008 and 0.005
magnitudes, respectively.

\section{Results}

\subsection {Isophotal Parameters}

The {\it final, dust-improved} (\S 2.3) $V$ and $I$ surface brightness
and $V-I$ color profiles, and the isophotal morphological parameters
derived from the isophotal fits to the (star-masked) $V$ and $I$
images, are given in Figure 1.  They will be made available in
computer format on request.  A comparison with the {\tt WF/PC} F555W
profiles of F95 and L95 is presented in Appendix B.

The third and fourth order sine and cosine terms indicate deviations
of the isophotes from a pure ellipse as described in terms of a
Fourier expansion.  In particular, a positive fourth order cosine term
indicates a disky structure, while a negative value for this
coefficient indicates boxy isophotes (e.g., Lauer 1985).

The color profiles systematically show scatter ($<0.05$ mag,
peak-to-peak) of a few points in the radial range from about 5'' to
9''. The inspection of the residual frames does not reveal any obvious
cause for this small scatter.  

\subsection{Individual Properties of the Galaxies}

The PSF-matched $V-I$ color maps for the 15 galaxies of our sample are
shown in Figure 2 (Plate XX). Each of the panels is $6''\times6''$ in
size. The frames are rotated so as to align horizontally the major
photometric axis of each galaxy.  Many galaxies possess a number of
features visible at small galactocentric radii.  In particular, as
already noted by F95, most of the objects show a significant amount of
patchy dust inside $1''$.  At the distances of our galaxies, the
detected features occur on scale lengths of the order of a few tens of
parsecs.  In the following we describe, separately for each galaxy,
the details of the isophotal properties derived from the fits to the
$V$ and $I$ frames, and the main features detected in them.

\paragraph{NGC 1427}

A visual inspection of the images shows a position angle twist inside
$\sim0.2''$, and very round isophotes inside this radius (as
indicated by the decrease in the ellipticity profile inside
$\sim0.3''$).  Although dust is not obvious in any of the (residual
and original) images, the $V-I$ color map shows two lanes, running
parallel to the minor axis and symmetric with respect to the
center. Their orientation is perpendicular to a red elongated
structure (inside $\sim1''$, oriented along the major axis). The
color map shows also some evidence for a redder color of the central
structure compared to surrounding regions.  The fit to the dust
improved  $I$ frame enhances the isophotal twisting.  The $C_4$ peaks
around $1''$, where it reaches the value of $\sim0.01$, and
decreases from there on, reaching slightly negative values outside
$10''$. The other Fourier terms are negligible over the range
$0.3''-15''$. These features all together indicate the presence of a
stellar disk between $0.3''-3''$ (which was suspected -- although not
detected -- by F95).

\paragraph{NGC 1439} 

The nuclear dust lane (at $\sim0.4''$) detected by F95 is clearly
identified with a ring in the color map.  SE of the nucleus
and adjacent to the ring, a red excess is detected (Figure 3).  The
ring is centered on the galaxy nucleus, has an apparent inclination of
$\sim50^\circ$ (assuming an intrinsic axial symmetry and $90^\circ
\equiv$ edge-on) and an extent of about $40 \times 65$pc.  The isophotal 
parameters are affected for radii less than the ring radius. 
There may also be some effect on the inner color profile.  From the ring
outward, the morphological parameters are unaffected by the dust, and
show a sharp decrease in ellipticity from about $1''$ to $2.5''$.  The
ellipticity peaks again at $\sim6''$, where also the $C_4$
suddenly rises to $\sim0.015$, and the isophotes show a small but
sudden change of position angle. This provides evidence for the
presence of a stellar ring at those radii.  Its average color is
similar to that of the nearby stellar population.

\paragraph{NGC 1700} 

Patchy dust and filaments are present in the innermost $2''$, and
influence the isophotal shape parameters inside $\sim 0.3''$.  The
patches are visible in the $V-I$ color map, which also shows a
spiral-like structure in the innermost $2''$ (200pc).  The galaxy is
boxy inside $3''$ ($C_4 \sim -0.01$), and disky outside this radius
($C_4 \sim 0.01$).  The ellipticity continuously rises 
outward, and shows structure clearly identified with the variations in
the fourth order cosine term.  The disky isophotes and the peak in
ellipticity indicate the presence of a stellar disk between $3''$ and
$10''$. No difference in color is seen between this disk and the
surrounding galaxy.

\paragraph{NGC 3608} 

The galaxy appears very round in the nucleus. The ellipticity
profile continuously rises with radius, albeit not very smoothly.
The $C_4$ shows a bump around $\sim 0.3''$.  
$S_3$ is also rather large at that radius, but is smaller in the 
fit to the dust-improved $I$ frame. 
This, and the slight decrease in ellipticity in that radial range,
suggests that the observed features are 
actually due to the presence of some patchy dust. This is
apparent from visual inspection of the residual $V$ image, and of the
$V-I$ color map, which  shows structure
inside the innermost 0.6$''$ (30 pc).
This feature appears elongated along the major axis,
and is reminiscent of a (weak) offcentered ring.
Outside $5''$, the galaxy is slightly boxy.

\paragraph{NGC 4278} 

One side of the galaxy is heavily obscured by large-scale dust located
NNW of the nucleus.  The dust distribution is far from settled,
and seems to spiral down into the nucleus. Its distribution shows
several dense knots, interconnected by filaments. The filaments are
roughly perpendicular to the extended radio emission. The densest and
largest of the knots appears about 5.4$''$ N of the nucleus, and is
$\sim$ 12 pc in size.   The large amount of dust dominates the 
higher-order terms of the Fourier expansion. These are reduced to
zero in the fit to the dust-improved $I$ frame, 
demonstrating the effectiveness of this procedure.  The color profiles
rise from the very center to $\sim1''$, and decrease toward bluer
values outside this radius.  A central point source has been found
($V\simeq19.7$, $M_V\simeq-9.6$, $V-I\simeq0.9$). It is bluer
than the surrounding stellar population (see the $V-I$ profile).
In addition, there is a weak indication from the $V-I$ color map 
that the stellar population surrounding the central point source is
also bluer than the surrounding galactic regions (note the slow 
increase in Figure 1 in $V_I$ to a radius of $\sim 1''$).

\paragraph{NGC 4365} 

The nucleus of this galaxy is very diffuse, i.e., the surface
brightness profile increases toward the center with a shallow
slope. However, it shows a weak (bump-like) excess of brightness in
the very center ($V \sim23$, $M_V\sim-7$) which might be unresolved.
Visual inspection of the $V$ image confirms the rounding of the
isophotes in the very center ($r\leq0.3''$) shown by the ellipticity
profile in this passband.  The $I$ image, instead, shows a weak
disk-like feature, which explains the larger ellipticity of the $I$
profile at those radii. A $\sim40^\circ$ isophote twist is also
observed there.  In our $V-I$ color map this central tilted structure
has a radius of about $0.14''$, and is bluer by about 0.05 mag than
the surrounding regions. It seems unlikely that this feature is an
artifact of an imperfect matching of the $V$ and $I$ PSFs, since its
size is a factor four larger than the FWHM of both Tinytim PSFs.  At
larger radii, a stellar disk (not seen by van den Bosch et al. 1994,
but detected by Forbes 1994) is clearly visible between $1''$ and
$3''$.  At these radii, the ellipticity and the $C_4$ show the typical
bump, where $C_4$ reaches 0.01.  There is a hint that this disk is
slightly redder than the outer surroundings.

\paragraph{NGC 4406} 

There is a slight decrease in the surface brightness in the 
region surrounding the very center of NGC 4406 (out to $\sim0.4''$ 
radius).   The shape of the obscuration is reminiscent of a nuclear
dust ring or torus seen close to face-on.  This ring appears extended
along the minor axis, i.e., its axis is aligned with the stellar
kinematic axis of the outer galaxy (which is tilted $90^\circ$ with
respect to that of the core). The ring is hardly seen in the $V-I$
color map, i.e., it does not give rise to a large amount of
reddening. However, it induces the large scatter  in the isophotal 
parameters inside $\sim 1''$ (since the scatter is reduced in the fit
to the dust-improved $I$ frame). Outside this radius, 
the isophotal parameters are well
defined, and show a transition between an internal stellar disk
($\sim 1''-6''$) and outer boxy isophotes.  The transition is
accompanied by a very weak ($\leq10^\circ$) isophote twist. No associated
structure is seen in the $V-I$ color map (although at those radii the
S/N might be inadequate).

\paragraph{NGC 4494} 

A very sharp dust ring (well-aligned with the apparent major axis, as was 
seen also by F95) is visible in the $V$ and $I$ images, and in the
$V-I$ color map.  The disk is centered with respect to the galaxy
center, but its intensity distribution is asymmetric, with the 
extinction minimum occurring at $\sim10$ pc from the center.  This is
very likely due to projection effects. In fact, its N and NW sides
produce a heavier reddening in the $V-I$ color map, and thus are
possibly closer to us. Assuming intrinsic circular geometry, the
apparent inclination of the ring is 60$^\circ$, and its projected size
roughly 90 pc $\times$ 50 pc.  The color map shows an asymmetry in the
region surrounding the disk, which, as in NGC 1439, is redder on the
far side of the ring.  The color profile shows a bump at the position
of the ring, and it is possibly influenced also at smaller radii.
Outside $\sim1''$ the isophotal parameters are not affected by the
dust ring. From that radius out to about $5''$, the ellipticity drops
rapidly from $\sim0.25$ to $\sim0.15$, and the $C_4$ from 0.05
to zero. A very small ($\sim10^\circ$) isophotal twist occurs in the
same region.  The regularity of the remaining Fourier terms confirms
the presence of a stellar disk inside $5''$.

\paragraph{NGC 4552}

The nucleus of this galaxy is obscured by an arc-shaped dust lane
which runs NE of it. The lane is clearly visible in the $V$ image and
in the $V-I$ color map, in which it resembles an incomplete (asymmetric) 
ring. More extended, radially elongated, filaments are also
visible in the frames.  Inside about 0.4$''$, the isophotal parameters
are strongly affected by the dust. Outside $\sim 1''$ the isophotes
become very round, a small isophotal twisting is present, and the higher
Fourier terms are negligible.  The $V-I$ color profile shows a bump
inside $\sim 0.25''$, most likely due to the nuclear dust.  The
central point source, reported as variable in the UV by Renzini et
al. (1995), is seen in our frames ($V \simeq 20.6$, $V-I \simeq 1.5$
mag). A detailed analysis of this and of the other two points
sources detected in NGC 4278 and IC 1459 will be reported elsewhere.

\paragraph{NGC 4589} 

Large scale patchy and filamentary dust perturbs the appearance of the
galaxy all across the PC chip.  The major dust lane runs almost N-S
(i.e., parallel to the minor axis) out to $\sim 10''$, and spatially
coincides with the position angle at which M{\o}llenhof \& Bender
(1989) observed the highest gas rotation velocity ($\pm 200$ km
s$^{-1}$; see also F95 and references therein).  This lane passes
close ($\sim0.1''$) to the apparent center, which is located in a
local minimum of extinction.  Around the nucleus and to its west, a
more irregular dust distribution is seen, including a cross-like feature
($\sim 9''$ away).  The large amount of dust affects the
morphological analysis, which shows a drop in ellipticity in the range
$0.1''-4''$, and relatively large values for the higher coefficients.
All these features are much reduced in the fit to the dust-improved
$I$ frame.

\paragraph{NGC 5322}

A completely obscuring dust lane, which is most probably a nearly
edge-on disk perpendicular to the broad radio-jet (Hummel et
al. 1984), cuts the nucleus in two, and hides the very center of
the galaxy. In the color map, the disk reaches a central value of
$V-I=1.62$, and remains on average about 0.25 mag redder than
the surroundings.  A red excess is detected in the region adjacent to
the disk, S of the galaxy.  The isophotal fitting within
$\sim2.5''$ is completely dominated by the strong dust absorption, 
as is clearly shown from the very large scatter in both the Fourier terms and 
the position angle.  Inspection of the residual maps shows that the sharp
decrease in ellipticity towards the center is also an artifact of the fits.  
Outside $2.5''$, the fits reproduce the morphology of the galaxy quite 
well.  The ellipticity drops significantly, together with $C_4$
which drops from $0.02$ to $-0.005$, indicating a transition from disky
to boxy isophotes. This provides strong  evidence for an embedded  stellar 
disk from $\sim1.5''$ out to $\sim$ $8''$.

\paragraph{NGC 5813}

A dust lane runs parallel to the major axis, East of the nucleus, and
contributes to producing the elongated appearance of the central
regions.  The extremely high values of ellipticity (and the scatter in
the Fourier coefficients) are an artifact of the dust lane, and are
partially reduced in the fit to the dust-improved $I$ frame.  Dust is
visible in the $V-I$ color map between the two apparent nuclei,
indicating that the double nucleus is an artifact of dust
obscuration.  Dust filaments are visible within $\sim7''$, and,
quite interestingly, are radially oriented toward the galaxy center. In
the $V-I$ color map, two of these filaments  are clearly visible
(oriented E and S,  respectively).  They are $\sim0.15$ mag
redder than the surrounding regions, and reach a $V-I$ value of 1.53
mag.

\paragraph{NGC 5982} 

The galaxy is extremely round within the innermost $1''$.  The scatter
of the other isophotal parameters in this region is larger than the
measurement errors, and is due to the shallow intensity slope.
At $\sim 1''$, a sharp increase in ellipticity and
decrease in $C_4$ is observed.  The galaxy remains flattened and boxy
($C_4\sim-0.02$) from $1''$ outwards. This behaviour might indicate
the presence of a face-on stellar disk which dominates the inner
region of an otherwise boxy galaxy. A second possibility, also
suggested by F95, is that it is an end-on bar that dominates the
optical light. Finally, it is also possible that the galaxy is
intrinsically round in the nucleus. The color profile is remarkably
flat at all radii; the $V-I$ color map shows an elongated feature E
of the nucleus, about 0.03 mag redder than the surrounding region.

\paragraph{NGC 7626}
 
A warped, symmetric dust feature (radius $0.45''$, seen also by F95)
crosses the very center of the galaxy, and heavily obscures it.  It
influences the isophotal parameters inside its radius. These show a
very steep increase in ellipticity which disappears in the fit to the
dust-improved $I$ frame.  The lane is tilted by $\sim 35^\circ$ with
respect to the radio jet (Jenkins 1982). A $20^\circ$ twisting between
$1''$ and $10''$ is observed.  This twist is barely significant, given
the low ellipticity of the isophotes.  The $V-I$ color map peaks at
the very center of the dust lane, with a value of 1.55 mag. The
entire dust lane is on average $\sim0.12$ mag redder than
the surroundings.  The color profile shows a continuous, significant
decrease with radius.

\paragraph{IC 1459} 

A heavy dust lane is present in the innermost $\sim 1.1''$, tilted
at $\sim15^\circ$ from the apparent major axis. The lane is
asymmetric with respect to the center, although it is present on both
sides of it. More chaotic patchy dust is present further away from the
nucleus, and other filamentary absorption features show up at
$\sim2''$ from the very center.  Inside the dust-lane region, a
sharp increase in ellipticity (whose amplitude is larger in $V$ than
in $I$) is observed, together with $\sim20^\circ$ twisting and a $C_4$
large and positive in the $I$ band.  However, at the same radii the $C_4$
values show a large scatter in the $V$ band, as do the other terms of the
Fourier expansion.  All these features are
much reduced in amplitude in the isophotal parameters derived from the
dust-improved $I$ frame.  We conclude that the major features are
induced by the dust rather than by real variations in the galactic
structure or in the stellar population, and that there is little photometric
evidence for the nuclear stellar disk suggested by the analysis of the
line-of-sight velocity distribution (Franx \& Illingworth 1988).
The $V-I$ color map traces the dust distribution, and
suggests the presence of a spiral-like
structure reminiscent of that detected on larger scales on deep
photographs (Malin 1985),  and also from the
ionized gas distribution (e.g., Forbes et al. 1990). The central point
source ($V\simeq18.4$, $M_V \simeq-12.7$ ) already detected by F95
stands out in the color map due to its $V-I$ color ($\sim 1.0$ mag),
which is bluer than the surrounding stellar population.

\subsection{Nuclear Colors and Color Gradients}

The nuclear colors derived with radial apertures of $0.5''$, $1''$,
$2''$ and $5''$ on the $V$ and $I$ images are reported in Table 2,
together with the logarithmic gradients $\frac {d (V-I)}{d Log r}$
between $0.25''\rightarrow 1.5''$ and $1.5''\rightarrow 10''$.  The
logarithmic color gradients should be rather robust against
contamination by patchy dust, since the $V$ and $I$ profiles were
obtained from isophotal fits to dust-masked frames (Sections 2.3 and
2.4).  Both the straight comparison of the different apertures and the
inner logarithmic gradients show that radial variations of $V-I$ exist
in the innermost $1.5''$.  However, the galaxies with the best
determined measurements (e.g., those with no contamination of point
sources or dust lanes) show inner gradients shallower than the outer
gradients (Figure 4).  For example, NGC 5982, one of the very few
galaxies of our sample without any obvious dust in the field of view
of the PC, shows a rather shallow color gradient inside $1.5''$.
Furthermore, galaxies with the highest central dust obscuration have
the steepest inner color gradients.  Another (apparently) dust-free
galaxy of our sample, NGC 4365, shows instead a reversal in its color
gradient.  The positive gradient of NGC 4365 arises from the central
blue region (about 15pc in size) detected in its color map. 

\section{Analytical Description of the Nuclear Properties}

\subsection{Parametrization of the Surface Brightness Profiles}

A few analytical expressions have been proposed to date for describing
the surface brightness profile in the inner regions of ellipticals
(e.g., F95; L95).  Here we chose to use the analytical representation
introduced by L95 (and extensively used by Byun et al. 1996), thereby
ensuring that a direct comparison could be made between the core
properties of ``normal'' ellipticals and those with
kinematically-distinct cores. This expression is:

\begin{equation}
\label{physlaw}
I(r) = 2^{(\beta-\gamma)/\alpha} I_b (\frac {r_b}{r})^\gamma \Big[ 1 +
( \frac {r}{r_b})^\alpha \Big]^{(\gamma-\beta)/\alpha}.
\end{equation}
The parameter  $\gamma$ measures the steepness of the rise of
the profile toward the very center (i.e., the value of $\gamma$ in
$I(r)\sim r^{-\gamma}$ as $r\rightarrow0$), $r_b$ is the break radius
where the profile flattens to a more shallow slope (in some cases),
$\beta$ is the slope of the outer profile, $\alpha$ controls the
sharpness of the transition between inner and outer profile, and $I_b$
is the surface brightness at $r_b$.  For the galaxies where a
continuous rise of the surface brightness profile is observed down to
the resolution of the data, the L95 law is underdetermined. However, it
still yields a very useful, smooth analytical representation of the
data. Given the undersampling of the {\tt WFPC2} PSF, we did not
perform any PSF-deconvolution, but chose instead to convolve the
models with the {\tt WFPC2} PSF before comparing them to the data.
The models also included a point source as necessary.

The fitting procedure was carried out in two steps. The first step
isolated the true minima of $\chi^2$ from secondary minima. The values
of the $\chi^2$ parameter were computed on a number of points
regularly distributed on a wide hypercube in parameter space (with
dimension equal to the number of free parameters), and once a minimum
value was found, a new, smaller, hypercube was placed on that
location, and the procedure iterated. The minimum value found on the
hypercube was then used as starting point to initialize a downhill
simplex minimization.  This procedure was tested on simulated data,
and found to recover the initial values with high accuracy.

We accepted as final the fits associated with the absolute minimum of
$\chi^2$. It is not possible to associate probabilities to these
$\chi^2$ values, since they are formally correct but not normalized.
There are many reasons for this. First, the data points we fit are not
statistically independent. Second, flat fielding errors and the
systematic mismatch, when present, between the theoretical law
[equation (4)] and the observed surface brightness profiles, conspire
so as to make it not trivial to give a strict statistical
interpretation of the $\chi^2$ values (see also Byun et al. 1996).  By
examining in detail the individual profiles, we found that in all
cases for which the formal $\chi^2\lta 1$, the fit is excellent, and
matches all the features present in the data.  The visual inspection
of the fits with $\chi^2\approx2$ ascertained that their quality was
slightly worse than that of the fits with smaller $\chi^2$.  None the
less, all our fits can be considered to be a good reproduction the light
profiles within the errors. The values of $r_b$, $\alpha$, $\beta$,
$\gamma$ and $\mu_b \equiv -2.5 Log_{10} I_b$ are listed in Table 3
for the fifteen galaxies of our sample.  For each of them, we used
these smooth representations of the data for deriving (i) the average
logarithmic nuclear slopes corrected for PSF and central point sources
effects (Section 4.2); (ii) the deprojected density profiles (Appendix
C); and (iii) the radius $r_{0.5}$ at which the slope of the best
fit surface brightness profile reached the value of 0.5. The latter
parameter is a more robust estimate of the bending radius than the
best fit value of $r_b$, which might be sensitive to
(e.g.,dust-induced) variations in the surface brightness profile. The
values of $r_{0.5}$ are listed in Table 4. There, the value $r_{0.5} <
0.1''$ has been assigned to all those galaxies whose profiles remained
steeper that $r^{-0.5}$ down to our resolution limit ($\sim0.1''$).
In order to compare the nuclear properties of kinematically-distinct
cores with those of kinematically-normal galaxies, we also derived the
$r_{0.5}$ values for the galaxies of the L95 sample (see Appendix
D).

\subsection{The Average Nuclear Logarithmic Slopes}

In order to provide a global representation of the nuclear galactic
properties, we computed average logarithmic slopes within a defined
radial range. This was taken to be $0.1''-0.5''$, i.e., coincident
with the range adopted by L95 for deriving an analogous quantity for
their sample of normal galaxies.  For both the $I$ and $V$ surface
brightness profiles, these logarithmic slopes $\langle
\gamma^{I,V}_{data} \rangle$ were initially computed fitting the data
points in two separate ways, i.e., through a least-squares fit and an
impartial fit.  The impartial fit combined the two least-square fits
obtained by exchanging the role of the dependent and independent
variables. This algorithm is generally statistically more robust than
the simple least-squares fit.  The two methods gave, within the
errors, identical results for the $\langle \gamma_{data}^{I,V}
\rangle$ values.

Since these fits to the data do not account for the
presence of central point sources or for the effect of the PSF, we
also computed the logarithmic slope $\langle \gamma_{fit}^{I,V}
\rangle$ (again within $0.1''-0.5''$) of the smooth theoretical
curves provided by the $V$ and $I$ best fits of equation
(\ref{physlaw}) to the data.  Good agreement was found between $\langle
\gamma_{fit} \rangle$ and $\langle \gamma_{data} \rangle$,
indicative of  the stability of the derived logarithmic slopes.  The
$\langle \gamma_{fit}^{I,V} \rangle$ values are listed in Table
5. As a reminder, a letter is given to identify those
nuclei in which dust is present in the form of a nuclear lane (L), of
an extended component (E), filaments (F), or of nuclear patches (P). A
question mark indicates galaxies where the detection of dust is
uncertain, while an asterisk indicates highly obscured nuclei (whose
best fit parameters are reported, but should be used with caution).

Compared to the asymptotic values of $\gamma$, the values of $\langle
\gamma_{fit} \rangle$ have the big advantage of being independent of 
any particular parametrization of the surface brightness profile. On
the other hand, they are more sensitive than $\gamma$ to distance
effects.  The more the surface brightness profile deviates from a pure
power-law, the more the $\langle \gamma_{fit} \rangle$ values vary
with distance (in galaxies with otherwise identical surface brightness
profiles). Thus, we also computed the logarithmic slopes $\langle
\gamma^{I,V}_{phys} \rangle$ between $10-50$ pc (an interval which
corresponds to about $0.1''-0.5''$ at the average distance of our
sample, i.e., $\sim 20 h^{-1} Mpc$). 

An inconvenience of both $\langle \gamma_{fit} \rangle$ and $\langle
\gamma_{phys} \rangle$ is that homologous galaxies differing only by a
scale factor would show very different values for these parameters.
However, since $\langle \gamma_{phys} \rangle$ does not depend on the
distance, it provides a valuable probe for exploring galactic
properties that would be subject to biases from distance
uncertainties.  A good correlation is found between the $\langle
\gamma_{fit} \rangle$ and $\langle \gamma_{phys} \rangle$ values,
which suggests that distance effects are not significant in our
sample.  This is not altogether surprising since the majority of our
galaxies typically lie within 10-20 $h^{-1}$ Mpc, and none lie beyond
40 $h^{-1}$ Mpc.  The $\langle \gamma^{I,V}_{phys} \rangle$ values,
both for the $V$ and $I$ profiles, are also given in Table 5.  In the
following, we adopt $\langle \gamma_{phys}^{I,V} \rangle$ to represent
the projected nuclear properties of the 15 galaxies.  As for the
$r_{0.5}$ values, we also derived the $\langle \gamma_{fit}^V \rangle$
and $\langle \gamma_{phys}^{V} \rangle$ values for the galaxies of the
L95 sample (see Appendix D).

\section {Discussion}

The discovery of kinematically-distinct cores 
gave support to the growing consensus that these smooth spheroidal
stellar systems might actually hide a past of violent mergers,
accretions and interactions with other galaxies. Given this colorful
picture, some signatures of such a hectic life might be expected in
the very centers of these galaxies.

\subsection{Properties of Kinematically-Distinct versus Normal Cores}

We have summarized the nuclear properties of our sample in Table 6.
The galaxies show a wide range in properties: most, but not all have
dust, many, but not all show disks, and the intensity profiles fall in
both catagories of ``shallow cusps'', and ``steep cusps''. There is no
unique, qualifying feature in the present sample. For 11 galaxies the
dust absorption is low enough that we can make a confident statement
about the presence of the stellar disk at a scale comparable to the
one of the kinematically-distinct nucleus (typically about $1''$). We
find that 7 out of the 11 galaxies show evidence for such a disk. Some
additional 
disks may have been missed because of projection effects (Rix \& White
1990). This suggests that such disks are very common in galaxies with
kinematically-distinct cores. It is difficult to state whether this is
a special properties of such galaxies, since there is no published
data set on ``kinematically-normal'' galaxies based on WFPC2 data.
This makes comparison of the relative disk-fractions in normal and
kinematically-peculiar galaxies impossible at this time.  However, it
appears that not all galaxies have a disk on the scale of the distinct
core. NGC 1700, for example, is slightly boxy on the scale of the
counter-rotating core.

The nuclear intensity profiles of kinematically-distinct cores show a very
large range of logarithmic slopes, from almost flat to very steep. The
same was observed by F95 with their smaller sample of such objects.
The presence of a break in the luminosity profiles is clearly not
directly related to the presence of a distinct component.
Furthermore, dynamical arguments (Merritt \& Fridman 1995) might exclude
the possibility that in some of these systems the observed decoupling
is due to projection effects of a triaxial figure (e.g., Statler
1991).  This, together with the fact that some of the
kinematically-distinct cores do not show minor axis rotation (which
rules out the models of Statler 1991), provides evidence that the
decoupling is often an intrinsic galactic property.

The comparison of the nuclear properties of kinematically-distinct
cores with those of galaxies that are (apparently)
kinematically-normal reveals a high degree of similarity between the
two classes of objects.  In Figure 5 the values of $\langle
\gamma_{phys}^V \rangle$ are plotted against the absolute $V$
magnitude M$_V$ for the galaxies of our sample (filled squares) and of
L95 sample (open triangles).  In Figure 6 the values of the break
radius $r_{0.5}$ are plotted against M$_V$, the symbols having the
same meaning as in Figure 5. The figures show that the average
logarithmic nuclear slopes and break radii of kinematically-peculiar
galaxies cover the same range of values spanned by galaxies with
normal core kinematics [although a hint is present for
kinematically-distinct core galaxies having, when compared to normal
galaxies, (i) intermediate values of $\langle \gamma_{phys}^V
\rangle$, and (ii) smaller values of $r_{0.5}$]. The universal
behaviour of nuclear slopes with luminosity was noted also by F95 in
their smaller sample.

L95 and K95 have pointed out that the  global and nuclear
properties are closely connected, with small galaxies hosting the
steepest nuclei.  This trend is confirmed in Figures 7 and 8, where
the values of $\langle \gamma_{phys}^V \rangle$ are plotted against
the central Mg$_2$ values of Davies et al. (1987) and the anisotropy
parameter $(v/\sigma_o)^*$ (expressed in logarithmic form, with $v$
the maximum rotation velocity and $\sigma_o$ the central velocity
dispersion of the galaxy; from Bender et al. 1992). The abscissas are
now essentially distance-independent, and galaxies leave a clear
pattern on these planes. Galaxies with steep cusps tend to be fast
rotators with a low $Mg_2$ absorption and large range of stellar
densities, while those with shallow cusps tend to be
supported by anisotropy in their velocity field, and to have strong
$Mg_2$ absorption and low stellar densities (see also Figure
9). Galaxies with kinematical peculiarities in their cores 
again closely follow the locus of normal galaxies.

It is a remarkable result that the nuclear morphology and structure of
galaxies with kinematically-distinct cores are so similar to
(apparently) kinematically-regular galaxies of the same luminosity
class. This result is even more puzzling when one considers the wide
dispersion in properties of the individual galaxies.  Consider, for
example, the cases of (i) NGC 1700, where the isophotes are very boxy
close to the galaxy center, and switch to a disky shape in a sharp
transition with increasing radius, (ii) NGC 4365, where the inner disk
smoothly dissolves into an outer boxy structure, and (iii) NGC 5982,
where the very round nucleus is embedded in a highly boxy galactic
body. These large changes in the shapes of the isophotes suggest that
distinct stellar components are present in the galactic potentials,
but that the nature of these components can vary quite considerably.
Nonetheless, the overlap of the nuclear properties of the
kinematically-peculiar galaxies with those of normal galaxies, and
their continuity of properties, suggests that the
kinematically-distinct cores may not owe their existence to any
particularly exotic event or process. The growth and evolution of the
nuclei appears to be closely related to the formation and evolution of
the entire galactic bodies, with the cores possibly being a later
development, or, alternatively, the kinematically-distinct cores are a
statistically possible outcome of universal processes which formed the
entire family of ellipticals, and their nuclei.

\subsection{Colors and Color Gradients of the Kinematically-Distinct Cores}

 The radial color profiles of galaxies with kinematically-distinct
cores provide a mean to trace variations in the stellar population
related to the possible existence of subcomponents.  Ground-based
photometric studies have not shown any anomalous color gradients for
these galaxies (see e.g., Franx et al. 1989, Peletier et al. 1990).
However, these studies, as do all ground-based studies, suffer from a
(seeing-) limited angular resolution.

In Figure 10 the $V-I$ color derived within an aperture of $5''$ is
plotted against the central Mg$_2$ value of Davies et al. (1987). The
spread in $V-I$ is large in our sample and might be attributable to
the large amount of dust detected in the nuclei. Some reduction was
found when the color was corrected for the average patchy extinction measured
within the same aperture on the $A_V$ and $A_I$ maps (as described in
\S 2.4). The correlation [$V-I = 0.76(\pm0.04)+1.70(\pm0.08) Mg_2$]
between the extinction-corrected color and the Mg$_2$ line strength
(which is unaffected by dust) is good, with a linear correlation
coefficient of 0.88.  This suggests that the nuclear colors provide a
reliable (if not unique) characterization of the properties of the
stellar populations.

With the exclusion of NGC 4365, the $V-I$ colors of the nuclear cusps
are rather normal.  By contrast, the central $\sim15$pc region of NGC
4365 contains an elongated structure bluer by $\sim0.05$ mag than the
surroundings. This blue feature might be due to, e.g., (i) a hole in
an otherwise uniform distribution of dust (a rather high dust mass has
been inferred for this galaxy, but we do not observe any obvious
patchy dust on the the PC image); or (ii) a pure stellar population
effect.  Assuming an underlying stellar population of 8 Gyrs and
[Fe/H]=0.25 dex, the difference in $V-I$ color between the nucleus and
its adjacent region implies either an age difference of 3-4 Gyrs, or a
metallicity difference of 0.1-0.2 dex in [Fe/H] (see models by Worthey
1994). The total luminosity of this component is $\sim 2.5 \times
10^6$ L$_\odot$, as derived from the total number of counts detected
in the $V$ image within an aperture comparable to the size of the
feature.  Observations of absorption line strengths at the proper
spatial resolution are required to confirm the nature of the blue
nucleus of NGC 4365, i.e., to confirm that differences in stellar
populations between the very nucleus and the surrounding regions might
exist in some kinematically-peculiar elliptical galaxies.

We stress that we have not detected any particular features in the
color profiles, which might be attributed to the
kinematically-distinct cores. Even in the seven galaxies in which
photometric evidence for a stellar disk (or ring) has been found,
namely NGC 1427, NGC 1439, NGC 1700, NGC 4365, NGC 4406, NGC 4494 and
NGC 5322, there is no evidence for any difference between the color of
the disk and that of the surrounding regions. The upper limit in the
$V-I$ color difference is about 0.02 mag.  
This difference should be contrasted to the color gradients at larger
radii, or the differences between galaxies, which span a range 
of 0.15 mag in our sample. It is therefore evident that the
stellar populations in the nuclei are surprisingly homogeneous.
Alternatively, a very finely tuned conspiracy between age variations
and metallicity variations much be present to produce very stable colors.
High resolution spectroscopy with HST is needed to confirm this
result. The spectroscopic indices will be much less sensitive to dust,
and can determine whether a ``conspiracy'' of changes exist to keep
the colors constant.

\subsection{Infall of a small galaxy and the role of a super massive
black hole}

The above results might be taken to rule out the scenario that the
kinematically distinct cores formed from a merger of a small
galaxy with a big galaxy (Kormendy 1984). In the original scenario,
the light at the nucleus was dominated by the small galaxy, and a
``core-within-a-core'' intensity profile was predicted.
The nuclear color was predicted to be blue. 
None  of these predictions have been confirmed, as noted by many
authors (e.g., Franx and Illingworth 1988).
Instead, these authors argued that the distinct components may have
formed in-situ, for example in a star burst. Schweizer (1989) argued
that full-fledged mergers might produce such structures.
One may therefore ask the question whether the
dissipationless merger hypothesis has been finally disproven.

It may be too early to do so.  For the galaxies with steep cusps, the
victim galaxy will be disrupted under most conditions, and no strong
population gradient would be expected anyhow.  For galaxies with
shallow cusps, the presence of super massive black holes may
substantial alter the fate of the small, infalling victim galaxy, as
argued earlier by F95. If the black hole has a mass comparable to the
mass of the shallow cusp, then the tidal force will remain high
throughout the whole shallow cusp.  We have performed simple,
semi-analytical calculations on the disruption of such a small victim
galaxy. These calculations are similar to the ones presented by F95,
and show that the light of the victim galaxy is distributed in a very
regular way over the host galaxy.  If M32 is allowed to spiral into IC
1459 and be disrupted by a super massive black hole, then the light
contribution would peak in the center at 10\%, and decrease in a very
regular way at larger radii. No strong color gradients would be
produced in this way.

Clearly, full simulations are needed to confirm the simple
calculations and to establish what kinematic signatures remain.  It is also clear, however, that it is too early to
regard models of dissipational infall as impossible on the basis of
the absence of color gradients.  To date, the strongest evidence
against these models remains the fact that we see nuclear disks being
formed in-situ in recent merger remnants.

\subsection{Evidence against Concentrated Diffused Dust}

Silva \& Wyse (1996) have suggested that a shallow nuclear cusp might
be observed if diffuse dust were concentrated in the nucleus, even
when the stars are actually distributed in a steep cusp.  Silva \&
Wyse's models predict a very steep $V-I$ color gradient associated to
the shallow cusp.

Excluding residual effects of the {\it patchy} dust on the color
profiles, strong $V-I$ gradients are not detected in our shallow-cusp
galaxies (see Figure 11, in which the inner $V-I$ gradients are
plotted against the nuclear slopes $\langle \gamma_{phys}^V
\rangle$). This is not what expected if dust were responsible for the
nuclear morphology and color gradients. We conclude that
diffuse dust is not the origin for the observed shallow cusps.

\subsection{Nuclear and Global Properties: Dichotomy or Continuum?}

Global and nuclear properties of elliptical galaxies are closely
related (e.g., L95, K95): {\it (i)} average- and low-luminosity
ellipticals rotate rapidly, are nearly isotropic, approximately
oblate-spheroidal, disky distorted, and (in the K95 terminology)
``coreless'' (i.e., have steep cusps), and {\it (ii)} giant
ellipticals, essentially do not rotate, are anisotropic, moderately
triaxial, boxy distorted, and have ``cuspy cores'' (i.e., have shallow
cusps). This implies that (small) galaxies with powerlaw cusps have
acquired or retained, in their luminous components, more angular
momentum than (large) galaxies with shallow cusps.  The distribution
of logarithmic nuclear slopes of the mass density profiles is bimodal,
i.e. it peaks around the values of -0.8 and -1.9 (Gebhardt et
al. 1996).  K95 proposed that the dichotomy in the observed properties
might actually imply a dichotomy in the formation processes that made
elliptical galaxies.

However, if our sample is added to that of the 'normal' L95 galaxies,
the trends of the nuclear slope versus the global physical parameters
are smooth and continuous, i.e. they do not show any sharp separation
between small and large, high- or low- angular momentum galaxies. For
example, {\it (i)} at intermediate luminosities (Figure 5) and
metallicities (Figure 7), some galaxies show steep stellar cusps,
while others have shallow cusps; {\it (ii)} although on average
anisotropic (pressure supported) galaxies have $\gamma_{phys}^V \lta
0.5$ and isotropic (rotation supported) galaxies have steeper nuclear
cusps, some degree of overlap is present (see e.g. Figure 12, where we
plot the nuclear cusp slope versus the dynamical mass scale $M \equiv
r_e \sigma_0^2 /G$, and distinguish the isotropic from the anisotropic
galaxies).

This leaves room to scenarios in which the observed 'dichotomy' in the
nuclear slopes actually arises from a {\it continuous variation} of
some parameter(s) governing the galaxy formation and evolution
processes. The observational results described so far suggest that one
of such parameters might be the amount of dissipation that occurred in
the course of galaxy formation.  Very likely, small, steep cusp
galaxies have suffered from a large amount of dissipation during their
formation. The shallow cusp galaxies might be instead the end products
of much less dissipative formation processes (e.g., White 1980;
Stiavelli \& Matteucci 1991).  More data are needed to resolve this
issue.

\section{Conclusions}

In this paper we have presented an analysis of {\tt WFPC2} F555W and
F814W images for fifteen galaxies with kinematically-distinct cores.
The color data were used to correct for the effects dust.  We have
derived surface brightness and isophotal parameter profiles in the two
bands, and color maps and radial profiles in $V-I$.  The surface
brightness profiles were parametrized by using an average logarithmic
slope inside 10-50 pc.  This was then used to investigate the nuclear
stellar properties of the galaxy cores.  A similar parametrization was
used for the nuclear stellar morphologies for the kinematically-normal
galaxies of L95, so as to compare the results obtained
for our sample with a reference sample.  We deprojected the surface
brightness profiles, and investigated the relationship between
projected quantities and physical stellar luminosity densities and
cusp slopes.

Our main results are:

\begin{itemize}

\item There is no unique photometric parameter that distinguishes the
galaxies with kinematically-distinct cores. They show a large range in
properties on the scale of the kinematically-distinct cores, like
their normal equivalents, namely shallow to steep cusps, and
small-scale disks and boxiness.  Many of the galaxies show dust, often
patchy and chaotic, and  sometimes distributed in a ring or disk.

\item With some possible exceptions, no obvious evidence is found for
homogeneous,  diffuse dust {\it confined} to the galactic nuclei.  Thus,
the nuclear colors are most likely representative of the stellar
population properties.

\item In one galaxy, namely NGC 4365, the innermost region (15pc in
size, for which the total luminosity is $\sim 2.5 \times 10^6$
L$_\odot$) is bluer by about 0.05 mag than the surrounding regions. If
this feature is interpreted as due to a variation in the stellar
population, synthetic population models suggest an age younger by
$\sim$ 3-4 Gyrs (or an $[Fe/H]$ variation of about -0.2 dex) between
the feature itself and the adjacent galactic regions.

\item The radial logarithmic slopes of the nuclear stellar profiles
are found to cover a large range of values, from almost flat to very
steep (the slopes $\gamma$ range from $\sim$ 0.1 to 0.8).  Some
galaxies show a clear ``break radius'', whereas other don't.  The
nuclear cusp slopes of kinematically-distinct core galaxies occupy the
same region of parameter space as that occupied by
kinematically-normal galaxies.  Independent of the core kinematical
properties, slowly-rotating, anisotropic, high metallicity galaxies
preferentially have shallow cusps, while fast-rotating,
low-metallicity galaxies tend to have steep cusps. Despite the bimodal
distribution of the cusp slopes, the trends of the cusp slope with
global galactic properties might be smooth and continuous, rather than
appearing as a distinct dichotomy, as indicated by e.g., K95.

\item Photometric evidence is found for faint stellar disks in seven
galaxies, namely NGC 1427, 1439, 1700, 4365, 4406, 4494 and 5322.
However, no difference in $V-I$ color ($<0.02$ mag) between these
disks and the surrounding galactic regions is present.  
These differences are much smaller than the gradients are larger
radii, or the color differences between galaxies (0.15 mag).
The nuclear populations are either very homogeneous, or finely-tuned
conspiracies between age and metallicity variations exist.
High resolution spectroscopy with HST is needed to test for the
existence of variations in the metal absorption lines.

\end{itemize}

These data have added  new constraints on aspects of the
formation process of elliptical and early-type galaxies.  While dust
has certainly complicated the interpretation of the data on very small
scales, techniques that minimize its effect can be used and allow one
to extract reliable quantitative information on structures, surface
brightness and luminosity density distributions, as well as on the
nature of the stellar population in the cores from colors.  Comparison
of the nuclear with the global properties suggests that
kinematically-distinct cores might be the normal outcome of regular
hierarchical merging at early epochs, and not indicative of any
unusual late event in the galaxies evolutionary history.

HST spectroscopy will be extremely valuable to study the nuclear
structure of the kinematically distinct components.  It remains
remarkable how these components stand out in the kinematics, and how
difficult they are to find in imaging.  HST spectroscopy may be able
to resolve many of the unanswered questions of this paper.

\acknowledgments

CMC thanks C. Norman, M. Stiavelli and T. de Zeeuw for useful
discussions; the Institute for Advanced Study, Princeton, and the
Space Telescope Science Institute, Baltimore, for hospitality.  CMC
was supported in the initial phases of this work at the Leiden
Observatory by HCM grant ERBCHBICT940967 of the European Community,
and is now supported by NASA through a Hubble Fellowship grant
HF-1079.01-96a awarded by the Space Telescope Institute, which is
operated by the Association of Universities for Research in Astronomy,
Inc., for NASA under contract NAS 5-26555.  This research was funded
in part by the HST grant GO-3551.01.-91A.

\bigskip
\bigskip

{\bf Appendix A: Deriving the patchy-dust improved  isophotal profiles}

\bigskip

The images of most galaxies showed evidence for patchy dust
absorption.
There can be no mistake that these features were due to dust.
They could always be characterized as irregular depressions in an
otherwise smooth intensity distribution, and they always corresponded
to red patches in the color maps.
We set out to obtain a simple correction of such absorption, to
produce ``dust-improved'' images.
We note that our procedure only ``corrects'' for the patchy dust
absorption, and not for any extended component.
Furthermore, it is not a full correction for the dust absorption:
 it gives only {\it improved} images.

The procedure is outlined below. It was applied to all galaxies
with evidence for patchy dust.

1. \ \ An initial model fit was made for the $I$ band, which is less
affected by dust than the $V$ band.  This model fit allowed the
position and the shape parameters (ellipticity and position angle) to
vary freely throughout the galaxy.  This procedure was re-iterated
after masking out spurious sources (e.g., foreground stars) on the
residual frame obtained with the initial model fit ('re-iterated' $I$
model).

2. \ \ A $V$ model was obtained while fixing the center and shape
parameters (ellipticity and position angle) of the $V$ isophotes to
those of the (re-iterated) $I$ model. The same mask as above was used
for the spurious sources. Dust-independent, intrinsic color gradients
might still be present, since at each radius the $V$ flux on the
isophote is a free parameter of the fitting procedure.

3. \ \  The assumption was then made that:
\begin{equation}
\frac {F^V_{obs}}{F^V_{intr}} = D + e^{-\tau}  (1-D)
\end{equation}
\begin{equation}
\frac {F^I_{obs}}{F^I_{intr}} = D + e^{-\alpha \tau} (1-D),
\end{equation}
where the subscripts {\it obs} and {\it intr} indicate the observed
and intrinsic (i.e., which would be observed in absence of dust)
emission, respectively, and $F^j$ indicates the counts in band $j$,
with $j=V,I$.  The free constant $D$ ranges between $0$ (screen in
front of the galaxy) and $1$ (screen behind the galaxy, not absorbing
any light), and $\alpha=\frac{A_I}{A_V}$ is the ratio between the
extinction in $I$ and in $V$. A value of $\alpha=0.5$ was assumed,
consistent with the Galactic extinction law (see, e.g., Rieke \&
Lebovsky 1985; Cardelli, Clayton and Mathis 1989). The two-dimensional
map of $\tau$ is the (unknown) optical depth as a function of
position.

In the equations above, the observed $V$ and $I$ images provide the
$F^V_{obs}$ and $F^I_{obs}$ maps, respectively, and the $V$ and $I$
models derived from the isophotal fits [steps (1.) and (2.)]  are assumed
to provide the $F^V_{intr}$ and $F^I_{intr}$ maps.

4. \ \ A two-dimensional map for the 'absorption' $a$ was defined as:
\begin{equation}
a \equiv \frac{F^V_{obs}/F^I_{obs}}{F^V_{intr} / F^I_{intr}}..
\end{equation}

5. \ \ By taking the ratio of the equations in step (3.), the absorption
$a$ was connected to the optical depth $\tau$:
\begin{equation}
a  = \frac {D+e^{-\tau} (1-D)}{D+e^{-\alpha \tau} (1-D)}
\end{equation}
The above equation can be turned into an algebraic equation for
$t=e^{-0.5\tau}$, namely:
\begin{equation}
t^2 - a t -\frac{D(a-1)}{1-D}=0
\end{equation}
This was solved to derive the map of $\tau$.

6. \ \ A 'dust-improved' $\Icorr$ frame was derived by using the
computed map of $\tau$ in the second equation of step (3.). 
\begin{equation}
F^\Icorr = [{D + e^{-\alpha \tau} (1-D)}]^{-1} F^I_{obs}.
\end{equation}
For clarity, we use here the subscript {\it improve} so as to avoid
confusion with the $I_{intr}$ map mentioned before [i.e. with the
're-iterated' $I$ isophotal fit model of step (1.)]. Although they are
defined by the same equation, $I_{intr}$ is an input frame when $\tau$
is unknown, while $\Icorr$ is an output frame once $\tau$ is
determined.

7. \ \ The isophote fitting procedure was applied to the
dust-improved $\Icorr$ frames. In the resulting morphological
profiles, most of the extreme features such as rapid variations of
ellipticity or of the position angle, and very high values for the
third and fourth order terms of the Fourier expansion of the intensity
(which indicate deviations of the isophotes from perfect ellipses),
were absent or much reduced in amplitude. 

The smoothness of the dust-improved $\Icorr$ profiles allowed us to
improve the determination of the center of ellipses, ellipticity and
position angle in the regions strongly affected by dust.

8. \ \ The final isophotal fits were performed, on the $I$ and $V$
images masked out of spurious point-sources and of dust, by keeping the
center of the ellipses, the ellipticity and the position angle fixed
to those derived from the fit to the $\Icorr$ image.  The dust mask
was produced by identifying all those pixels strongly obscured by dust
on the residual frames derived from the model fits of steps (1.) and
(2.).

9. \ \ The sequence outlined in steps (3.) to (8.) was repeated for
several values for 'screen-location' parameter $D$. The value of $D=0$
already provided smooth $I$ maps, devoid of any obvious patchy dust,
i.e. already provided smooth profiles for the center of ellipses, the
ellipticity and the position angle needed in step (8.).  Values of $D$
different from zero (i.e. screens 'inside' the galaxy) did not
significantly affect the resulting isophotal morphological
parameters. Thus, in our successive analysis, we decided to adopt the
$\Icorr$ frame obtained by setting $D=0$.

\smallskip

The azimuthally averaged surface brightness profiles resulting from
this procedure are termed {\it final, dust-improved} in our text.
They are a much improved representation of the underlying galaxy
starlight.  Typical errors on the light profiles, estimated from the
residuals of the model fits, are $\sim0.02$ mag.

\bigskip
\bigskip

{\bf Appendix B: Comparison with {\tt WF/PC} F555W Profiles}

\bigskip

F95 and L95 presented F555W profiles for some of the galaxies of our
sample. In particular, IC 1459, NGC 1427, 1439, 4365, 4494, 4589, 5982
and 7626 were investigated by F95, while NGC 1700, 3608 and 5813 are
included in L95.  In Figure 13 we compare our F555W
surface brightness, ellipticity, and position angle profiles within
$0.1''-2.5''$ to those derived from those authors from the {\tt WF/PC}
data.  The agreement is generally good, once NGC 1700 and NGC 4589
are excluded from the comparison. For the former, L95 published a
profile oriented differently, by about 90$^\circ$, and 
for the latter no isophotal
parameters were reported by F95 due to the extensive dust. For the
surface brightness profiles, the three galaxies of L95 are in very
good agreement with our measurements (once an average zero-point shift of about
$0.05$ mag is removed).  Also the surface brightness profiles
published by F95 for IC 1459, NGC 4365 and NGC 5982 match our
profiles quite well.  For the remaining galaxies (namely, NGC 1427, NGC 1439, NGC
4494, NGC 4589 and NGC 7626), a systematic discrepancy is found in the
innermost $\sim0.3''$, where the F95 profiles are brighter than the
{\tt WFPC2} ones (by $\Delta \mu \lta 0.4$ mag).  Possible causes
for this effect are: {\it (i)} an imperfect correction for the {\tt
WF/PC} PSF via the Richardson-Lucy method in galaxies with steep
central cusps. In this respect, it is remarkable that L95 recovered
the light profiles in the inner $0.4''$ with a Fourier algorithm,
rather than via the standard deconvolution procedure; {\it (ii)} the
use of different dust masks in the obscured nuclei; and {\it
(iii)} the influence of the residual HST PSF on the {\tt WFPC2}
profiles.  These factors are all potentially problematic, 
especially on strongly-cusped nuclei.

\bigskip
\bigskip

{\bf Appendix C: The Physical Density Profiles}

\bigskip

The observed surface brightness profiles were deprojected by means 
of an Abel inversion (assuming spherical symmetry) in order to
derive the physical luminous density $\rho$ (as a function of radius).
There are problems associated with the simple-minded use of such an inversion.
When used directly on the data this procedure amplifies the fluctuations and errors
present in the data. In addition, since the inner profiles are
influenced by the PSF, the deprojection of the data is also affected
by resolution effects.  These effects are minimized by using the 
smooth galaxy models obtained by fitting equation (\ref{physlaw}) 
to the observed profiles,
since the PSF was automatically taken into account in our fitting
procedure (with the potential risk here of systematic effects if the model does
not match extremely well).  Thus, in order to obtain an estimate of the reliability of
the derived values, we deprojected both the data and the smooth
theoretical models.  These two deprojected profiles
were mostly in very good agreement.  Both deprojections were finally fitted
with equation \ref{physlaw} in order to have a parametric
representation of the luminous density.  For each galaxy, the
asymptotic ($r\rightarrow0$) slope $\gamma_\rho$ of the fit to the
deprojection of the model surface brightness was adopted as
representative of the intrinsic slope of the nuclear cusp.

A quality parameter was defined for the deprojections, ranging from
zero (least reliable), to 3 (most reliable).  This was obtained by
averaging between the two approaches the number of times any of the
following properties was satisfied: (1) absence of a central
point-like source; (2) steepness of the inner surface brightness
profile ($\langle \gamma_{fit} \rangle$) smaller than 0.5; and (3) a
difference smaller than 0.1 between {\it (a)} the asymptotic slope
$\gamma_\rho$ derived from the deprojection of the smooth theoretical
surface brightness profile, and {\it (b)} the equivalent quantity
computed from the original data.

The quality parameter is listed in Table 7 together
with the $V$ luminous density at $0.1''$ and 10 pc in $L_\odot/pc^3$,
the deprojected logarithmic slopes in the range $0.1''-0.5''$ and
$10-50$ pc ($\gamma_{\rho,0.1-0.5''}^{V,I}$ and
$\gamma_{\rho,10-50pc}^{V,I}$, respectively), and the asymptotic
values of $\gamma_\rho^{V,I}$.  In the table, all the given values are
those derived from the deprojections of the model surface brightness,
while the errors are the absolute difference between these
estimates and those derived from direct of the data.

In Figure 14 the average logarithmic slope of the $V$
surface brightness profiles $\langle \gamma_{phys}^V \rangle$ is
plotted against the corresponding asymptotic slope $\gamma_\rho$ (for
ours and the L95 sample; see Appendix C).  As expected (see e.g., K95),
the points lie below the line $\gamma_\rho = \langle \gamma_{phys}^V
\rangle + 1$; however, the correlation is good. Thus, the relationships of
the galactic properties with $\langle \gamma_{phys}^V \rangle$ can be
directly translated into relationships with $\gamma_\rho$, although
the former was adopted in our discussion because of its direct link to
the observed data.

\bigskip
\bigskip

{\bf Appendix D: Nuclear Cusp Slopes and Densities for the L95 Sample}

\bigskip

L95 derived from their $V$ surface brightness profiles a parameter
$\gamma_{L95}$, defined as the ``average slope over the interval
$0.1''-0.5''$ calculated from the total change in brightness implied
by the fitting law (equ. \ref{physlaw}) across the radial interval''.
These values of $ \gamma_{L95}$ should in principle be directly
comparable to our $\langle \gamma_{fit}^V \rangle$ values.  However,
in order to guarantee an unbiased comparison, we judged it safer to
derive the $\langle \gamma_{fit}^V \rangle$ values also for the
galaxies of L95, starting from their published profiles.  For the
three galaxies NGC 1700, 3608 and 5813, both the {\tt WF/PC} data of
L95 and our {\tt WFPC2} data were available.  The values of $\langle
\gamma_{fit}^V \rangle $ derived for these galaxies from the two
samples are in very good agreement. The largest difference occurs for
NGC 1700, the steepest of the three galaxies, and is equal to 0.08.
As expected, the agreement between $\gamma_{L95}$ and $\langle
\gamma_{fit} \rangle$ is always very good (Figure 15).

In addition to $\langle \gamma_{fit}^V \rangle $, also the values of
$\langle \gamma_{phys}^V \rangle $ and of the break radius $r_{0.5}$ were
derived following the same definition adopted for the galaxies of our
sample. In Table 8 these measurements are reported,
together with the estimates of the break radii published by L95.   Those 
galaxies that are unresolved are indicated by $r_{0.5}<0.1''$.
  
 Fits to the surface brightness profiles of the L95 sample were
deprojected as in Appendix C, to allow direct comparison with our
galaxies.  The derived physical densities at $1''$ are compared to
those published by L95 in Figure 16.  Similarly, Figure 17 shows the
asymptotic deprojected slopes obtained with our procedure plotted
against the values published by K95 (panel a) and Gebhard et
al. (1996; panel b). The values published by K95 and Gebhardt et
al. (1996) were obtained by means of a non-parametric deprojection.
The independent sets of measurements show a good agreement.  This
strengthens our confidence in the reliability of the deprojected
parameters.

In our comparisons with normal galaxies, we excluded NGC 1700, 3608
and 5813 from the L95 sample. Furthermore, we considered only the L95
galaxies covering the same range of distance as our own sample
($D\leq40$Mpc).

\newpage

\small

\begin{table*}
{
\small
\begin{tabular}{llrrrrrrllllr}
&&&&&&&& \\
\hline\hline
Name     & Type & D     & $R_e$&  M$_V$   &B-V$_o$&Mg$_2$&X-ray   & A$_V$ \\
\hline								 
NGC 1427 & E5   & 13.19 & 2.53 &  -19.77  &  0.93 &0.258 &-       &0.0    \\
NGC 1439 & E1   & 15.50 & 3.13 &  -19.81  &  0.93 &0.269 &-       &0.05   \\
NGC 1700 & E4   & 40.55 & 2.72 &  -21.71  &  0.93 &-     &-       &0.09   \\
NGC 3608 & E2   & 10.20 & 1.76 &  -19.29  &  0.98 &0.312 &39.19   & 0.0   \\
NGC 4278 & E1   &  7.30 & 1.17 &  -19.26  &  0.96 &0.291 &-       & 0.08  \\
NGC 4365 & E3   & 10.35 & 2.82 &  -20.42  &  0.99 &0.321 &39.48   & 0.0   \\
NGC 4406 & E3   & 10.35 & 4.50 &  -21.09  &  0.89 &0.311 &40.75   & 0.08  \\
NGC 4494 & E1   & 11.20 & 2.49 &  -20.46  &  0.90 &0.275 &-       & 0.05  \\
NGC 4552 & E0	& 10.35 & 1.05 &  -20.20  &  0.97 &0.324 & ?      & 0.11  \\
NGC 4589 & E2   & 18.30 & 3.71 &  -20.66  &  0.94 &-     &39.88   & 0.03  \\
NGC 5322 & E3   & 21.05 & 3.63 &  -21.42  &  0.89 &0.276 &39.93   & 0.0   \\
NGC 5813 & E1   & 15.93 & 4.39 &  -20.56  &  0.94 &0.308 &-       & 0.11  \\
NGC 5982 & E3   & 29.92 & 2.07 &  -21.35  &  0.95 &0.296 &40.45   & 0.02  \\
NGC 7626 & E1P  & 37.15 & 6.85 &  -21.86  &  1.00 &0.336 &40.82   & 0.12  \\
IC  1459 & E3   & 16.55 & 3.13 &  -21.20  &  0.99 &0.321 &40.45   &0.0    \\
\hline
&&&&&&&& \\
\end{tabular}
}
\caption{
Parameters for the 15 galaxies of the sample.
Column 1: IC/NGC number;
Column 2: Morphological Type (from RC2);
Column 3: Distance ($h^{-1}$ Mpc; from Faber et al. 1989, or Bender et al. 1992);
Column 4: Effective radius ($h^{-1}$ kpc; from Faber et al. 1989, or 
Bender et al. 1992);
Column 5: Absolute $V$ magnitude (mag + 5 log $h$;
Faber et al. 1989, or Bender et al. 1992);
Column 6: Corrected $B-V$ color (magnitudes; from 
Faber et al. 1989, or Bender et al. 1992);
Column 7: Central Mg$_2$ index  (magnitudes; from Davies et al. 1987);
Column 8: Log[L$_x$] ($h^{-2}$ erg s$^{-1}$;
from Roberts et al. 1991);
Column 9: Galactic extinction in $V$ (magnitudes; from Faber et al. 1989,
after assuming A$_V$=0.75A$_B$. The extinction in $I$ was 
derived assuming A$_I$=0.51A$_V$).
}
\label{sample}
\end{table*}

\begin{table}
\begin{center}
\begin{tabular}{lrrrrrr}
\hline\hline
\multicolumn{1}{l}{Name} &
\multicolumn{4}{c}{$V-I$}&
\multicolumn{2}{c}{$d (V-I)/ d log r$} \\
\multicolumn{1}{l}{} &
\multicolumn{1}{c}{$0.5''$} &
\multicolumn{1}{c}{$1.0''$} &
\multicolumn{1}{c}{$2.0''$} &
\multicolumn{1}{c}{$5.0''$} &
\multicolumn{1}{c}{$0.25''-1.5''$} &
\multicolumn{1}{c}{$1.5''-10''$} \\
\hline
N1427 & 1.26 & 1.25 & 1.24 & 1.22 &     $-0.056 \pm0.002$ &             $ -0.104\pm0.035$\\
N1439 & 1.35 & 1.32 & 1.29 & 1.26 &     $-0.199 \pm0.009$ &             $ -0.100\pm0.040$\\
N1700 & 1.36 & 1.35 & 1.33 & 1.31 &     $-0.073 \pm0.002$ &             $ -0.087\pm0.012$\\
N3608 & 1.30 & 1.30 & 1.28 & 1.27 &     $-0.048 \pm0.001$ &             $ -0.106\pm0.029$\\
N4278 & 1.31 & 1.32 & 1.33 & 1.33 &     $0.049  \pm0.003$ &             $ -0.079\pm0.019$\\
N4365 & 1.32 & 1.32 & 1.32 & 1.31 &     $0.022  \pm0.011$ &             $ -0.053\pm0.027$\\
N4406 & 1.31 & 1.30 & 1.30 & 1.29 &     $-0.022 \pm0.002$ &             $ -0.064\pm0.005$\\
N4494 & 1.36 & 1.31 & 1.27 & 1.23 &     $-0.237 \pm0.005$ &              $-0.075\pm0.019$\\
N4552 & 1.37 & 1.36 & 1.35 & 1.33 &	$-0.046 \pm0.001$ &	      $ -0.072\pm0.004$\\
N4589 & 1.46 & 1.44 & 1.42 & 1.35 &     $-0.114  \pm0.002$&         $     -0.211\pm0.014$\\
N5322 & 1.39 & 1.33 & 1.29 & 1.25 &     $-0.160  \pm0.001$&          $    -0.048\pm0.011$\\
N5813 & 1.41 & 1.39 & 1.38 & 1.36 &     $-0.056  \pm0.003$&           $   -0.116\pm0.025$\\
N5982 & 1.27 & 1.27 & 1.26 & 1.26 &     $-0.025 \pm0.003$ &            $  -0.083\pm0.042$\\
N7626 & 1.43 & 1.41 & 1.40 & 1.37 &     $-0.072 \pm0.002$ &             $ -0.100\pm0.020$\\
I1459 & 1.43 & 1.37 & 1.36 & 1.33 &     $-0.104  \pm0.001$&             $ -0.084\pm0.007$\\
\hline
\end{tabular}
\caption{$V-I$ colors in magnitudes, and logarithmic color gradients
for the 15 galaxies.  Column 1: Galaxy name.  Column 2: $V-I$ color in
the $0.5''$ aperture.  Column 3: $V-I$ color in the $1.0''$ aperture.
Column 4: $V-I$ color in the $2.0''$ aperture.  Column 5: $V-I$ color
in the $5.0''$ aperture.  Column 6: $V-I$ color gradients in the range
$0.25''-1.5''$.  Column 7: $V-I$ color gradients in the range
$1.5''-10''$.  For IC 1459, NGC 4278 and NGC 4552, the point sources
were subtracted before performing the measurements. Internal errors
for the aperture measurements are smaller than 0.01 magnitudes; for
the gradients, the formal errors of the fits are reported. }
\label{VIgradientstab}
\end{center}
\end{table}

\begin{table*}
\begin{center}
{\tiny
\begin{tabular}{llllllll}
\hline\hline
\multicolumn{1}{c}{Name} &
\multicolumn{1}{c}{Filter} &
\multicolumn{1}{c} {$r_b$} &
\multicolumn{1}{c} {$\alpha$} &
\multicolumn{1}{c}{$\beta$} &
\multicolumn{1}{c}{$\gamma$} &
\multicolumn{1}{c}{$\mu_b$} &
\multicolumn{1}{c}{$\chi^2$} \\
\hline
N1427 & V & 0.897 &1.817 &1.339 &0.680 & 16.501 &0.55\\
N1427 & I & 0.815 &1.760 &1.349 &0.674 & 15.157 &0.44\\
N1439 & V & 0.495 &23.61 &1.326 &0.777 & 16.027 &1.65\\
N1439 & I & 0.424 &20.03 &1.338 &0.776 & 14.570 &1.95\\
N1700$^*$ & V & 0.181 &0.462 &1.649 &0.007 & 14.582 &0.14\\
N1700$^*$ & I & 0.179 &0.475 &1.676 &0.007 & 13.273 &0.23\\
N3608 & V & 0.453 &0.725 &1.582 &0.003 & 15.782 &1.08\\
N3608 & I & 0.418 &0.780 &1.566 &0.003 & 14.424 &0.87\\
N4278 & V & 0.894 &1.452 &1.316 &0.    & 15.982 &0.89\\
N4278 & I & 1.112 &1.252 &1.464 &0.    & 14.850 &2.30\\
N4365 & V & 1.979 &1.669 &1.460 &0.108 & 16.949 &0.85\\
N4365 & I & 1.826 &1.516 &1.489 &0.045 & 15.582 &0.59\\
N4406 & V & 0.945 &4.132 &1.046 &0.041 & 16.082 &0.20\\
N4406 & I & 0.912 &3.322 &1.075 &0.    & 14.809 &0.21\\
N4494 & V & 0.840 &4.125 &1.232 &0.613 & 15.802 &1.78\\
N4494 & I & 0.700 &3.250 &1.246 &0.604 & 14.394 &1.73\\
N4552$^*$ & V & 0.478 &2.174 &1.056 &0.    & 15.160 &0.37\\
N4552$^*$ & I & 0.497 &2.095 &1.083 &0.043 & 13.886 &0.39\\
N4589 & V & 0.757 &0.433 &1.622 &0.    & 16.608 &1.51\\
N4589 & I & 0.541 &0.501 &1.578 &0.009 & 14.911 &0.74\\
N5322 & V & 0.705 &1.235 &1.401 &0.    & 15.875 &0.39\\
N5322 & I & 0.683 &1.107 &1.452 &0.    & 14.586 &1.10\\
N5813 & V & 0.889 &1.767 &1.414 &0.029 & 16.560 &0.69 \\
N5813 & I & 0.872 &1.667 &1.456 &0.008 & 15.232 &0.53\\
N5982$^*$ & V & 0.476 &2.155 &1.188 &0.118 & 15.890 &0.17\\
N5982$^*$ & I & 0.472 &2.165 &1.192 &0.117 & 14.655 &0.22\\
N7626 & V & 0.406 &1.530 &1.225 &0.001 & 16.250 &1.13\\
N7626 & I & 0.390 &1.295 &1.274 &0.    & 14.910 &1.30\\
I1459 & V & 1.433 &0.959 &1.757 &0.016 & 16.248 &2.42\\
I1459 & I & 1.822 &0.807 &2.036 &0.020 & 15.148 &1.21\\
\hline
\end{tabular}							         
}								
\caption{Structural parameters from the fits of equation (4) to the
$V$ and $I$ surface brightness profiles of our galaxies. For some of
the galaxies (indicated by an asterisk), the fits had to be restricted
within $3''$, in order to better reproduce the nuclear profiles. For
the remaining galaxies, the fits were extended to $10''$.}
\label{apertdusttab}
\end{center}
\end{table*}

\begin{table}\begin{center}
\begin{tabular}{lll}
\hline\hline
\multicolumn{1}{l}{Name} &
\multicolumn{1}{l}{$r^V_{0.5}$} &
\multicolumn{1}{l}{$r^I_{0.5}$} \\
\hline
N1427 & $<0.1$ & $<0.1$ \\
N1439 & $<0.1$ & $<0.1$ \\
N1700 & $<0.1$ & $<0.1$ \\
N3608& $0.15\pm 0.07$ & $0.16\pm 0.08$\\
N4278& $0.64\pm 0.10$ & $0.66\pm 0.10$\\
N4365& $1.16\pm 0.08$ & $1.09\pm 0.03$\\
N4406& $0.91\pm 0.10$ & $0.87\pm 0.10$\\
N4494 & $<0.1$ & $<0.1$ \\
N4552& $0.46\pm 0.02$ & $0.44\pm 0.01$\\
N4589& $0.12\pm 0.04$ & $0.11\pm 0.32$\\
N5322& $0.44\pm 0.19$ & $0.38\pm 0.13$\\
N5813& $0.61\pm 0.10$ & $0.59\pm 0.10$\\
N5982& $0.36\pm 0.18$ & $0.36\pm 0.18$\\
N7626& $0.32\pm 0.20$ & $0.28\pm 0.05$\\
I1459& $0.53\pm 0.10$ & $0.43\pm 0.10$\\
\hline
\end{tabular}\end{center}
\caption{Values of the break radius $r_{0.5}$ in arcseconds derived from
the best fits to the $V$ and $I$ surface brightness profiles as the
points at which the (model) nuclear profiles become shallower than
$r^{-0.5}$. Galaxies in which this condition was never reached are
considered unresolved ($r_{0.5} \leq 0.1''$).  Errors are the differences
between the $r_{0.5}$ derived from the best model fits, and those obtained
directly from the data. In order to obtain the latter values, the
point source contribution was subtracted when present, and the data
smoothed when too noisy. }
\label{rb05}
\end{table}

\begin{table}
\begin{center}
\begin{tabular}{lcrrrr}
\hline\hline
\multicolumn{1}{l}{Name} &
\multicolumn{1}{l}{Dust}&
\multicolumn{1}{l}{$\gamma^V_{fit}$} &
\multicolumn{1}{l}{$\gamma^I_{fit}$} &
\multicolumn{1}{l}{$\gamma^V_{phys}$} &
\multicolumn{1}{l}{$\gamma^I_{phys}$} \\
\hline
N1427& ${L,?}$&   $0.74\pm0.02$&  $0.75\pm0.02$  &       $0.80\pm0.04$   &       $0.82\pm0.03$   \\
N1439& $L$&       $0.79\pm0.03$&  $0.83\pm0.03$  &       $0.86\pm0.10$   &       $0.93\pm0.10$   \\
N1700& $P$&       $0.86\pm0.02$&  $0.87\pm0.01$  &       $0.68\pm0.08$   &       $0.65\pm0.09$   \\
N3608& $P$&       $0.59\pm0.04$&  $0.59\pm0.04$  &       $0.80\pm0.02$   &       $0.82\pm0.03$   \\
N4278& ${E,F,*}$& $0.20\pm0.05$&  $0.21\pm0.07$  &       $0.53\pm0.04$   &       $0.51\pm0.04$   \\
N4365&   &        $0.15\pm0.02$&  $0.12\pm0.02$  &       $0.24\pm0.05$   &       $0.23\pm0.04$   \\
N4406& ${L,?}$&   $0.05\pm0.04$&  $0.03\pm0.02$  &       $0.16\pm0.02$   &       $0.18\pm0.04$   \\
N4494& $L$&       $0.61\pm0.04$&  $0.63\pm0.01$  &       $0.67\pm0.04$   &       $0.71\pm0.01$   \\
N4552& $L,F$ &    $0.23\pm0.06$&  $0.25\pm0.06$  &       $0.53\pm0.04$   &       $0.54\pm0.04$   \\
N4589& ${E,F,*}$& $0.59\pm0.01$&  $0.61\pm0.01$  &       $0.60\pm0.03$   &       $0.62\pm0.01$   \\
N5322& ${L,*}$&   $0.30\pm0.02$&  $0.34\pm0.01$  &       $0.29\pm0.05$   &       $0.33\pm0.04$   \\
N5813& ${F,P}$&   $0.17\pm0.01$&  $0.18\pm0.02$  &       $0.24\pm0.03$   &       $0.24\pm0.02$   \\
N5982&     &      $0.34\pm0.01$&  $0.34\pm0.01$  &       $0.21\pm0.02$   &       $0.21\pm0.03$   \\
N7626& ${L,*}$&   $0.38\pm0.03$&  $0.43\pm0.01$  &       $0.18\pm0.03$   &       $0.22\pm0.01$   \\
I1459& ${L,*}$&   $0.34\pm0.04$&  $0.36\pm0.05$  &       $0.36\pm0.05$   &       $0.41\pm0.06$   \\
\hline			   
\end{tabular}		   
\caption{$I$ and $V$ logarithmic slopes within $0.1''-0.5''$
($\gamma_{fit}$) and 10-50pc ($\gamma_{phys}$) for the galaxies of the
sample.  Column 1: IC/NGC number; Column 2: Dust morphology: a nuclear
lane (L), an extended component (E), filaments (F), or nuclear patches
(P). A question mark indicates galaxies where the detection of dust is
uncertain. An asterisk indicates an heavily obscured nucleus.  Column
3: Logarithmic slopes of the $V$ surface brightness profiles within
$0.1''-0.5''$; Column 4: Logarithmic slopes of the $I$ surface
brightness profiles within $0.1''-0.5''$; Column 5: Logarithmic slopes
of the $V$ surface brightness profiles within 10-50pc; Column 6:
Logarithmic slopes of the $I$ surface brightness profiles within
10-50pc.  Tabulated values are the logarithmic slopes of the best model fits
to the data; errors are the differences between these estimates, and those
obtained through a straight fit to the data.}
\label{gamma12tab}
\end{center}
\end{table}

\tiny
\begin{table}
\begin{center}
\begin{tabular}{lcrrrr}
\hline\hline
\multicolumn{1}{l}{Name} &
\multicolumn{1}{l}{Comments}&
\multicolumn{1}{l}{Stellar Disk?} &
\multicolumn{1}{l}{Cusp} \\
\hline
N1427  &dust inside $r \simeq 0.1''$; disky inside $1''$, boxy outside & Y& steep \\
N1439  &dust ring at $r \simeq 0.4''$; $\epsilon$ peak at $1''$  & Y& steep   \\
N1700  &patchy dust; boxy inside $r\simeq 1$'', disky outside	   & Y&  steep   \\
N3608  &no significant structure                           & N& shallow \\
N4278  &extensive dust; blue point source                       & N&  shallow$^*$ \\
N4365  &$\epsilon$ peak inside $1-5''$; disky inside $\simeq2''$, boxy outside; blue nucleus& Y &  shallow\\
N4406  &dust ring; disky inside $1-5''$, boxy outside &	Y  & shallow  \\
N4494  &strong dust ring; $\epsilon$ high near $1''$; slightly disky &Y& steep\\
N4552  &patchy dust inside $r \simeq 0.4''$; filamentary dust outside; point source &       N &  shallow \\
N4589  &extensive dust	&					? & shallow cusp$^*$ \\
N5322  &strong dust disk; $\epsilon$ peak inside $2.5$-$5''$; disky	& Y & shallow$^*$ \\
N5813  &patchy/filamentary dust; no other structure	&	N  & shallow \\
N5982  &strong isophotal twist and low $\epsilon$ inside $r \simeq 1''$ & Bar? & shallow \\
N7626  &small dust warp; no other structure		&	N  & shallow  \\
I1459  &dust lane; somewhat disky outside $r \simeq 1''$; blue point source &	? & shallow \\
\hline
\end{tabular}
\caption{ Nuclear properties of the 15 galaxies of our sample. Column
2: summary of dust and isophotal morphology; Column 3: Y and N
indicate the presence or absence of a stellar disk, respectively;
Column 4: shape of the intensity profile at the scale of the distinct
nucleus.  Question marks indicate uncertain detections; asterisks in
Column 4 indicate uncertain measurements due to strong dust
obscuration.}
\label{marijn}
\end{center}
\end{table}

\begin{table*}
{\tiny \begin{center}
\begin{tabular}{llrrrrrr}
\hline\hline
\multicolumn{1}{l}{Name} &
\multicolumn{1}{l}{Filter} &
\multicolumn{1}{r}{$\gamma_{\rho,0.1-0.5"}$} &
\multicolumn{1}{r}{$\gamma_{\rho,10-50pc}$} &
\multicolumn{1}{r}{$\rho^V_{0.1''}$} &
\multicolumn{1}{r}{$\rho^V_{10pc}$ } &
\multicolumn{1}{r}{$\gamma_{\rho}$}&
\multicolumn{1}{r}{Quality}\\
\hline 
N1427 &V&$1.63\pm 0.001$ &$1.64\pm 0.06$ &$1765\pm 20$ &$890 \pm 20$ &$1.38\pm 0.22$&1.5 \\
N1427 &I&$1.62\pm 0.04$ &$1.65\pm 0.02$ & & &$1.32\pm 0.10$&\\
N1439 &V&$1.63\pm 0.06$ &$1.57\pm 0.03$ &$1620\pm 120$  &$1035 \pm 50$ & $0.98\pm 0.29$ &1.5\\
N1439 &I&$1.58\pm 0.07$ &$1.60\pm 0.12$ & & &$1.35\pm 0.09$& \\
N1700 &V&$1.74\pm 0.10$ &$1.58\pm 0.33$ &$1000\pm140 $ &$2800\pm 1200$ &$1.08\pm 0.19$&1.5 \\
N1700 &I&$1.76\pm 0.13$ &$1.59\pm 0.48$ & & &$0.82\pm0.06$& \\
N3608 &V&$1.34\pm 0.21$ &$1.58\pm 0.33$ &$1380\pm 360$ &$635\pm 55$ &$0.73\pm 0.17$&1\\
N3608 &I&$1.33\pm 0.22$ &$1.59\pm 0.48$ & & &$0.84\pm 0.46$& \\
N4278 &V&$0.47\pm 0.34$ &$1.03\pm 0.04$ & $410 \pm 400$ & $295 \pm 40$ &$0.17 \pm 0.55$& 1 \\
N4278 &I&$0.54\pm 0.07$ &$1.03\pm 0.09$ & & &$0.24\pm 0.23$& \\
N4365 &V&$0.69\pm 0.33$ &$0.67 \pm 0.08$ &$175\pm 55$ &$105\pm5$ &$0.55\pm 0.01$&2.5 \\
N4365 &I&$0.47\pm 0.25$ &$0.57\pm 0.01$ & & &$0.35\pm 0.18$& \\
N4406 &V&$0.26\pm 0.54$ &$0.14\pm 0.45$ &$115\pm 45$ &$85\pm 35$ &$0.18\pm 0.18$ & 2.5  \\
N4406 &I&$0.00\pm 0.40$&$0.13\pm 0.44$ & & &$0.01\pm 0.01$&  \\
N4494 &V&$1.52\pm 0.01$ &$1.44\pm 0.12$ &$2650\pm 200$ &$1060\pm 30$ &$1.48\pm 0.16$&1 \\
N4494 &I&$1.48\pm 0.12$ &$1.48\pm 0.11$ & & &$1.42\pm 0.24$& \\
N4552 &V&$0.41\pm 0.14$ &$0.96\pm0.02$ & $480\pm80$ & $448\pm 96$ & $0.02\pm0.43$ & 1.5 \\
N4552 &I& $0.54\pm 0.17$ & $1.01\pm 0.04$ & & & $0.26 \pm 0.06$ & \\
N4589 &V&$1.43\pm 0.13$ &$1.45\pm 0.09$ &$600\pm 30$ &$535\pm 95$ &$1.04\pm 1.00$&1 \\
N4589 &I&$1.44\pm 0.23$ &$1.46\pm 0.18$ & & &$0.92\pm 0.92$& \\
N5322$^*$ &V&$0.73$ &0.71 &$260\pm 130$ &$265\pm 90$ & $0.25$&2.5 \\
N5322$^*$ &I&$0.85$ &$0.83$ & & &$0.03\pm 0.02$&  \\
N5813 &V&$0.42\pm 0.07$ &$0.49\pm 0.02$ &$110\pm 20$ &$105\pm 10$ &$0.44\pm 0.28$&2.5 \\
N5813 &I&$0.40\pm 0.10$ &$0.49\pm0.05$ & & &$0.13\pm 0.03$&  \\
N5982 &V&$0.74\pm 0.12$ &$0.63\pm 0.23$ &$1720\pm 30$ &$225\pm 70$ &$0.56\pm 0.38$& 2 \\
N5982 &I&$0.74\pm 0.13$ &$0.62\pm 0.23$ & & &$0.56\pm 0.51$& \\
N7626 &V&$0.83\pm 0.17$ &$0.47$ &$100\pm 35$ &$115\pm 115$ &$0.15$&2 \\
N7626 &I&$0.98\pm 0.17$ &$0.64$ & & &$0.58$& \\
I1459 &V&$0.79\pm 0.40$ &$0.85\pm 0.05$ & $415\pm 555$ & $370\pm 260$ &$0.72\pm 0.05$ & 2\\
I1459 &I&$0.93\pm 0.03$&$0.99\pm 0.13$ & & & $0.96\pm 0.08$ & \\
\hline
\end{tabular}\end{center}
}
\caption{Deprojected average logarithmic and asymptotic slopes, and
luminous densities (in $L/pc^3$), for the galaxies of our sample.
Column 1: Name and filter; Column 2: Logarithmic slopes within
$0.1''-0.5''$; Column 3: Logarithmic slopes within $10-50$pc; Column
4: Luminous $V$ densities at $0.1''$; Column 5: Luminous $V$ densities
at $10$pc; Column 6: Asymptotic slope derived from fitting the
analytical law (Eq. 4) to the deprojections of the surface brightness
profiles; Column 7: Quality parameter. Errors are the difference
between the estimates obtained from deprojecting the model surface
brightness profiles, and those obtained from deprojecting the
data. Errors are not reported for those cases where the straight
deprojection of the data could not be performed. ($^*$) The measurements
for NGC 5332 are uncertain, due to the heavy dust obscuration of its
nucleus. }
\label{intrinsictab}
\end{table*}

\tiny
\begin{table}\begin{center}
\begin{tabular}{lrrr}
\hline\hline
\multicolumn{1}{l}{Name} &
\multicolumn{1}{l}{$\langle \gamma^V_{phys} \rangle$} &
\multicolumn{1}{l}{$r^V_{0.5,our}$} &
\multicolumn{1}{l}{$r^V_{b,L95}$} \\
\hline
N524  &  $0.52\pm0.04  $  &$0.30\pm0.03$ & 0.33\\
N596  &	 $0.82\pm0.04  $  &$<$0.1& $<$0.08\\
N720  &	 $0.07\pm0.01  $  &$2.44\pm0.09$ & 3.21\\ 
N1023 &	 $0.74\pm0.01  $  &$<$0.1&$<$0.05\\ 
N1172 &	 $1.00\pm0.06  $  &$<$0.1&$<$0.05\\ 
N1331 &	 $0.59\pm0.18  $  &$<$0.1&$<$0.05\\ 
N1399 &	 $0.13\pm0.01  $  &$1.73\pm0.22$ & 3.14\\ 
N1400 &	 $0.62\pm0.01  $  &$0.18\pm0.01$ & 0.33\\ 
N1426 &	 $0.84\pm0.03  $  &$<$0.1&$<$0.05\\ 
N2636 &	 $0.95\pm0.07  $  &$<$0.1&$<$0.1\\ 
N2841 &	 $0.76\pm0.01  $  &$0.15\pm0.09$ & $<$0.1\\ 
N3115 &	 $0.85\pm0.05  $  &$<$0.1&$<$0.05\\ 
N3377 &	 $1.15\pm0.04  $   &$<$0.1&$<$0.08\\ 
N3599 &	 $0.95\pm0.06  $  &$<$0.1&$<$0.05\\ 
N3605 &	 $0.69\pm0.01  $  &$<$0.1&$<$0.08\\ 
N4239 &	 $0.41\pm0.01  $  &$0.31\pm0.59$ & $<$0.05\\ 
N4387 &	 $0.72\pm0.01  $  &$<$0.1&$<$0.08\\ 
N4434 &	 $0.86\pm0.03  $  &$<$0.1&$<$0.05\\ 
N4458 &	 $1.40\pm0.08  $   &$<$0.1&$<$0.1\\ 
N4464 &	 $0.97\pm0.05  $  &$<$0.1&$<$0.05\\ 
N4467 &	 $0.95\pm0.05  $  &$<$0.1&$<$0.05\\ 
N4486B&	 $1.16\pm0.04  $   &$0.11\pm0.03$ & 0.18\\ 
N4551 &	 $0.81\pm0.02  $  &$<$0.1&$<$0.05\\ 
N4636 &	 $0.19\pm0.02  $  &$2.04\pm0.41$ & 3.21\\ 
N4697 &	 $0.67\pm0.01  $  &$<$0.1&$<$0.05\\ 
N4742 &	 $1.02\pm0.01  $   &$<$0.1&$<$0.05\\ 
N5845 &	 $0.51\pm0.06  $  &$0.20\pm0.15$ & $<$0.05\\ 
N7332 &	 $1.00\pm0.03  $  &$<$0.1&$<$0.05\\ 
V1199 &	 $1.22\pm0.21  $   &$<$0.1&$<$0.05\\ 
V1440 &	 $0.99\pm0.13  $  &$<$0.1&$<$0.05\\ 
V1545 &	 $0.62\pm0.04  $  &$<$0.1&$<$0.05\\ 
V1627 &	 $1.09\pm0.17  $   &$<$0.1&$<$0.05\\ 
\hline
\end{tabular}\end{center}
\caption{Values of the average logarithmic slope within 10-50pc
($\langle \gamma_{phys}^V \rangle$), and of the break radius in
arcseconds, for the galaxies of the L95 sample within 40 $h^{-1}$ Mpc.  Column
2: Tabulated values of $\langle \gamma_{phys}^V \rangle$ are the
logarithmic slopes of the best model fits to the data; errors are the
differences between these estimates, and those obtained through a
straight fit to the data.  Column 3: our measurements of $r_{0.5}$,
derived from the best fits to the $V$ surface brightness profiles as
the points at which the nuclear profiles become shallower than
$r^{-0.5}$. In our measurements, we consider unresolved those nuclei
in which $r_{0.5} <0.1''$.  Errors are the differences between the $r_{0.5}$
derived from the best model fits, and those obtained directly from the
data.  Column 4: the measurements of $r_b$ published by L95.  }
\label{rblauer}
\end{table}

\newpage

{\bf Figure 1.} Surface brightness and isophotal parameters as a
function of log(radius), in arcsec.  Stars are used for the $V$ band,
open squares for the $I$ band.  From top to bottom, the left panels
show the surface brightness $\mu$, the $V-I$ color, the position angle
($PA$) and the ellipticity ($\epsilon$) profiles; the right panels the
third order sine ($S_3$) and cosine ($C_3$), and fourth order sine
($S_4$) and cosine ($C_4$) terms.  In the $V-I$ color panels, dots
indicate the original unconvolved profiles; crosses indicate the color
profiles obtained after convolution to ensure that the $V$ and $I$
data have the same PSF (see \S 2.4).  The thick solid line indicates
the fit to the dust-corrected $I_{improve}$ frames. Formal errors from
the ellipse fitting are given for the isophotal parameters. For the
surface brightness profiles, errors include sky subtraction. Errors on
the $V-I$ values are the quadratic mean of the errors on the $V$ and
$I$ bands. The errors are given every three points; when they are not
visible, they are smaller than the size of the symbols.

\bigskip

{\bf Figure 2.} $V-I$ color maps for the 15 galaxies. Each panel is
$6''\times6''$.  The photometric major axes are aligned with the
abscissa. Redder regions are darker.  From left to right, first row:
NGC 1427, NGC 1439, NGC 1700, NGC 3608; second row: NGC 4278, NGC
4365, NGC 4406, NGC 4494; third row: NGC 4552, NGC 4589, NGC 5322, NGC
5813; last row: NGC 5982, NGC 7626, IC 1459.

\bigskip

{\bf Figure 3.} Three cuts across the $V-I$ color map of NGC 1439.
Each cut is the average within a strip 7 pixels wide, running parallel to
the major axis. The solid line passes through the galaxy center; the dotted line 
is to the N, and the dashed to the S.
The sharp dust ring is symmetric with respect to the galaxy center.
An area of weaker reddening surrounds the
ring. This area is more extended on the far side of the ring.

\bigskip

{\bf Figure 4.} Comparison of inner ($0.25''-1.5''$) and outer
($1.5''-10''$)  $V-I$ color gradients for the galaxies of our sample.
The galaxies with the best determined measurements are plotted
as filled squares.

\bigskip

{\bf Figure 5.} Projected nuclear logarithmic slope $\langle
\gamma_{phys}^V \rangle $ versus absolute magnitude $M_V$. Filled
squares are our kinematically-distinct core galaxies, open triangles
the ``normal'' L95 galaxies.

\bigskip

{\bf Figure 6.} Logarithm of break radius $r_{0.5}$ (in pc) versus
absolute magnitude $M_V$. Filled squares are our sample, open
triangles the L95 galaxies.  Galaxies with values of $r_{0.5} < 0.1''$
are not included in the plot.  The solid line shows the value of
$r_{0.5}$ in pc corresponding to $0.1''$ at the closest distance of
our sample (7.3 $h^{-1}$ Mpc). The arrow indicates the effect of an
error of 50 \% in the distance.
  
\bigskip

{\bf Figure 7.} Projected nuclear logarithmic slope $\langle \gamma_{phys}^V
\rangle $  versus
central Mg$_2$ values (in magnitudes; from Davies et al. 1987).
Filled squares are our sample,  open triangles the
L95 galaxies. 

\bigskip

{\bf Figure 8.} Projected nuclear logarithmic slope $\langle \gamma_{phys}^V
\rangle $ versus values of the
anisotropy parameters $(v/\sigma_\circ)^*$ (from Bender et al. 1992).
Filled squares are our sample,  open triangles the
L95 galaxies. 

\bigskip

{\bf Figure 9.} Values of stellar densities at 10pc, in
$L_\odot/pc^3$, versus central Mg$_2$ (in magnitudes; from Davies et
al. 1987). Squares are our sample, triangles the L95 galaxies. Open
symbols are for galaxies with $\langle \gamma_{phys} ^V \rangle <
0.5$, filled symbols for galaxies with steeper slopes.

\bigskip

{\bf Figure 10.} Measurements of $V-I$ inside $5''$ versus the Mg$_2$
values of Davies et al. (1987; filled triangles).  Units are
magnitudes along both axes. Internal errors on the aperture
measurements of $V-I$ are negligible. The 6-point stars are obtained
by correcting the $V-I$ measurements for reddening (using the values
of $E(V-I)_{aperture}$, see \S 2.4).  The good correlation supports
that the nuclear $V-I$ is not strongly affected by dust, and is thus a
reliable indicator of the properties of the stellar populations in the
central cusps.

\bigskip

{\bf Figure 11.}  Logarithmic $V-I$ color gradients inside
$0.25''$-$1.5''$ versus $\langle \gamma_{phys} ^V \rangle$. Identified
are the galaxies with heavy patchy dust, or dust lanes.

\bigskip

{\bf Figure 12.}  Dynamical mass scale $log M = log (\sigma_o^2 r_e
/G)$, in $M_\odot$, versus the nuclear logarithmic cusp slope $\langle
\gamma^V_{phys} \rangle$.  The squares are our sample of
kinematically-distinct core galaxies, and the triangles are the
galaxies of L95.  Filled symbols are anisotropic galaxies with $log
(v/\sigma_o)^* <-0.3$, open symbols are isotropic rotators, where $log
(v/\sigma_o)^* > -0.3$. The values of $\sigma_o$, $r_e$ and
$log(v/\sigma_o)^*$ are from Bender et al. (1992).

\bigskip

{\bf Figure 13.} Comparison with the {\tt WF/PC} F555W data of F95 and L95.
Each panel contains, from top to bottom, the position angle PA, the
ellipticity $\epsilon$, and the $V$ surface brightness profiles.  First
row, from left to right: NGC 1427, NGC 1439, NGC 1700, NGC 3608;
second row, left to right: NGC 4365, NGC 4494, NGC 4589, NGC 5813;
third row, left to right: NGC 5982, NGC 7626, IC 1459. 

\bigskip

{\bf Figure 14.} Luminosity density, $V$ band, deprojected asymptotic logarithmic slopes
$\gamma_\rho$ versus projected average nuclear slope $\langle
\gamma_{phys} ^V \rangle$. The $\gamma_\rho =
\langle\gamma_{phys} ^V\rangle+1$ line is plotted for reference.

\bigskip

{\bf Figure 15.} Comparison between the logarithmic slopes 
$\gamma_{L95} $ and $\langle \gamma_{fit} \rangle$ for the L95 sample (see text).

\bigskip

{\bf Figure 16.} Luminous density at $0.1''$ (in $L_\odot/pc^3$) derived by
L95 and from this work.  
  
\bigskip

{\bf Figure 17.} Panel (a): Deprojected asymptotic logarithmic slopes
 derived by Kormendy et al.  (1995) and from this work for the six
 galaxies of the L95 sample with shallowest nuclear slope. The filled
 symbol represents NGC 6166, for which the errorbars are very large
 along both axes. Panel (b): Deprojected asymptotic logarithmic slopes
 derived by Gebhardt et al.  (1996) and from this work for those
 galaxies for which both methods provided reliable slopes.

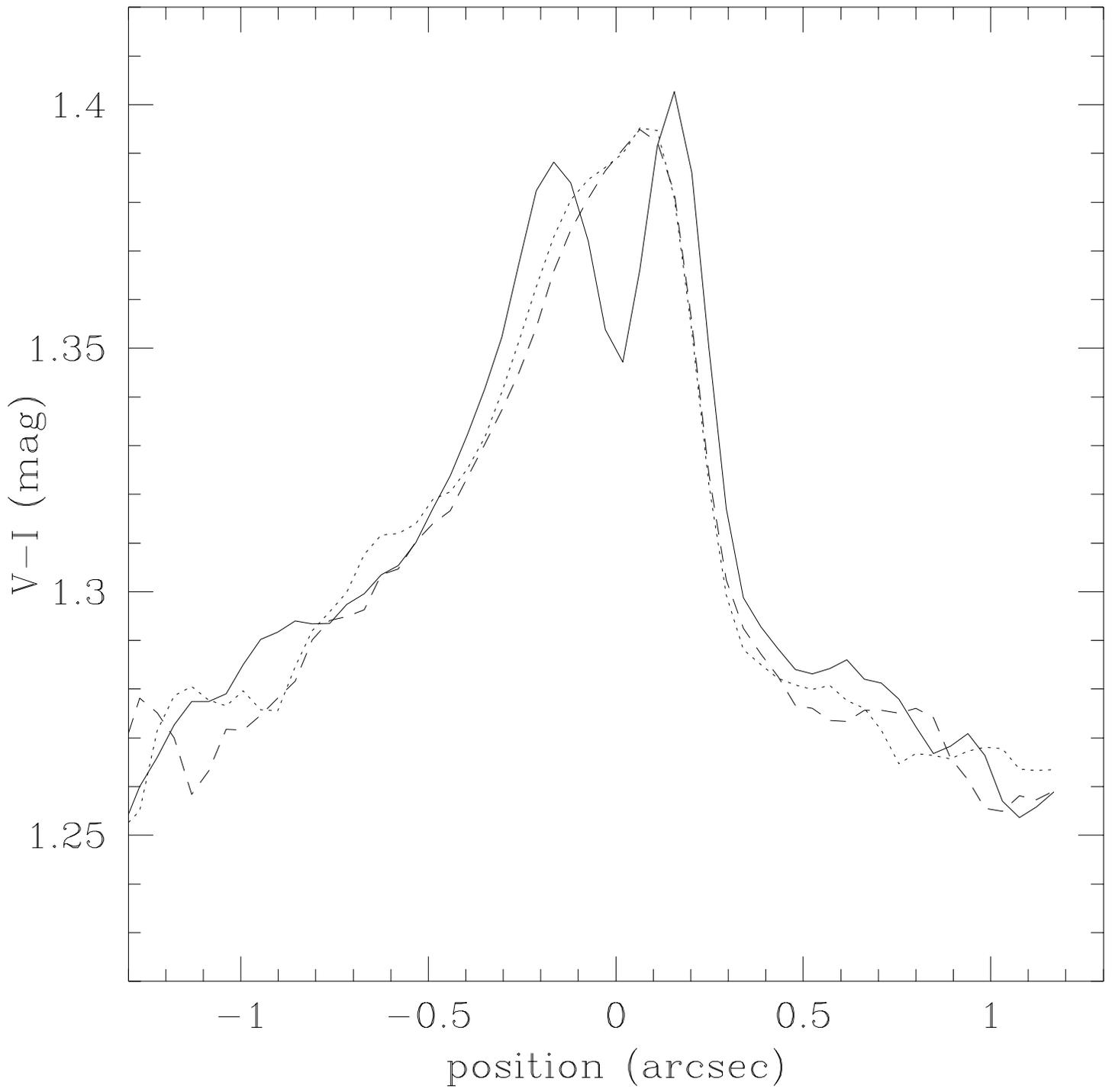
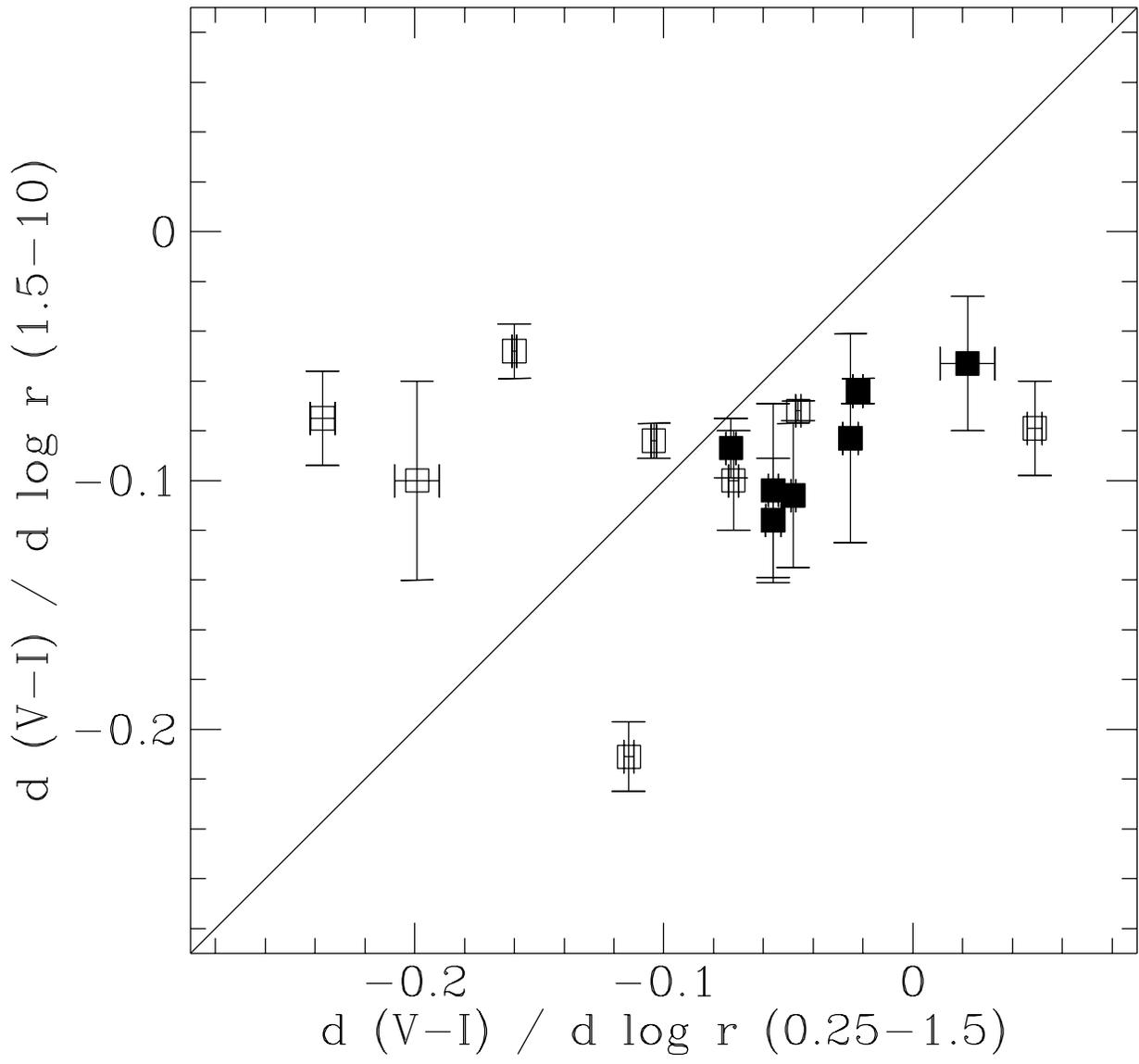
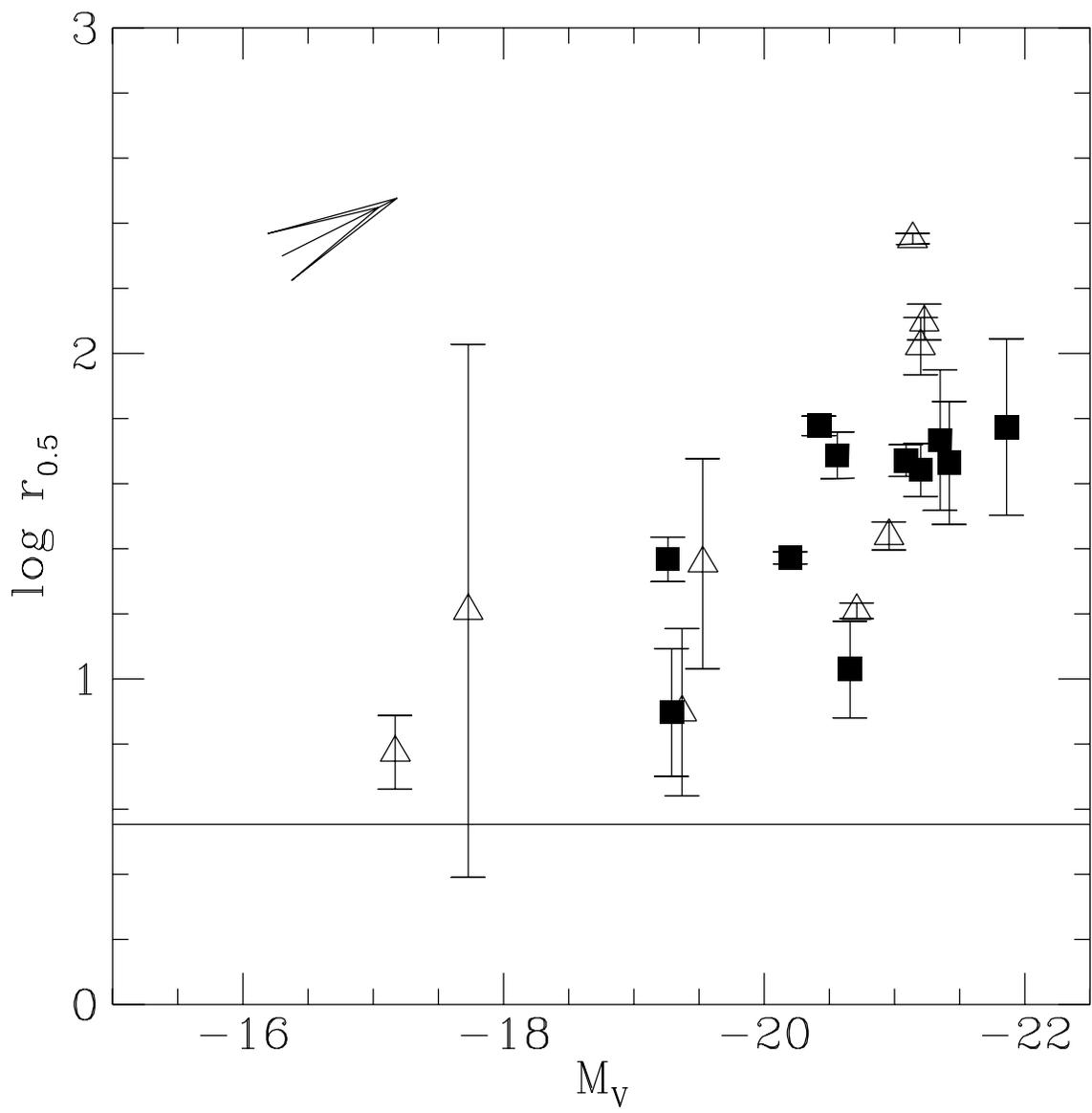
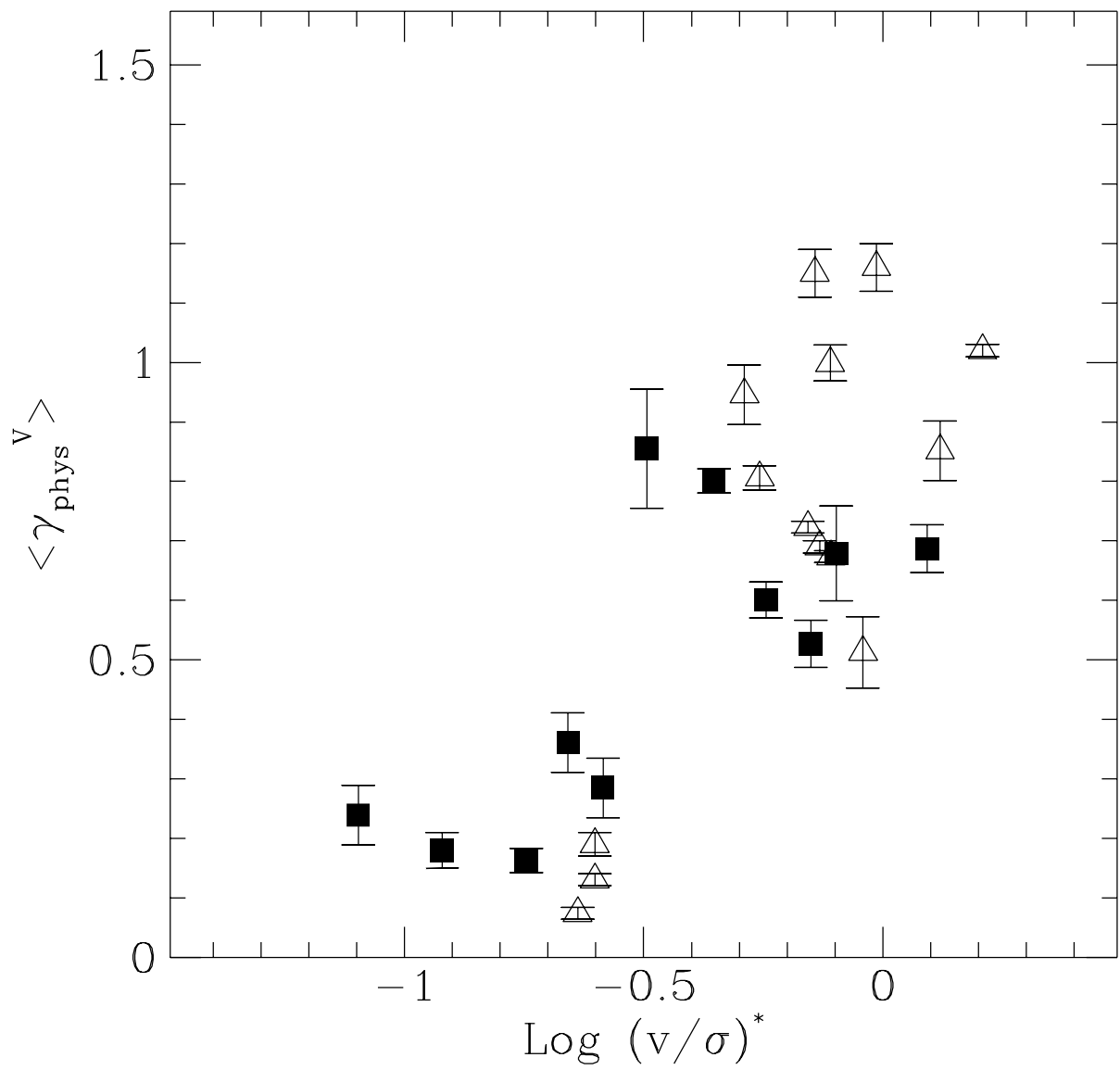
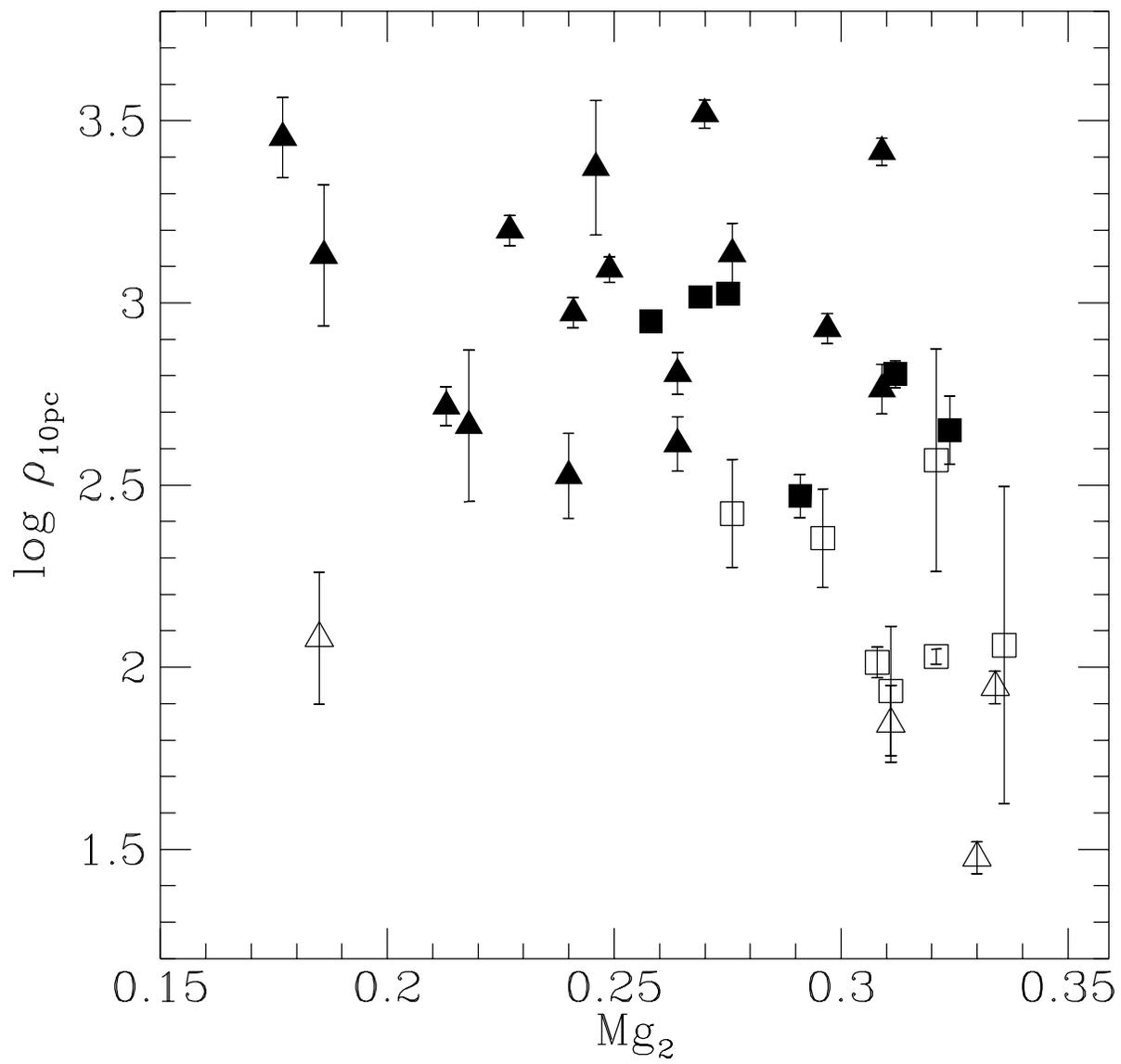
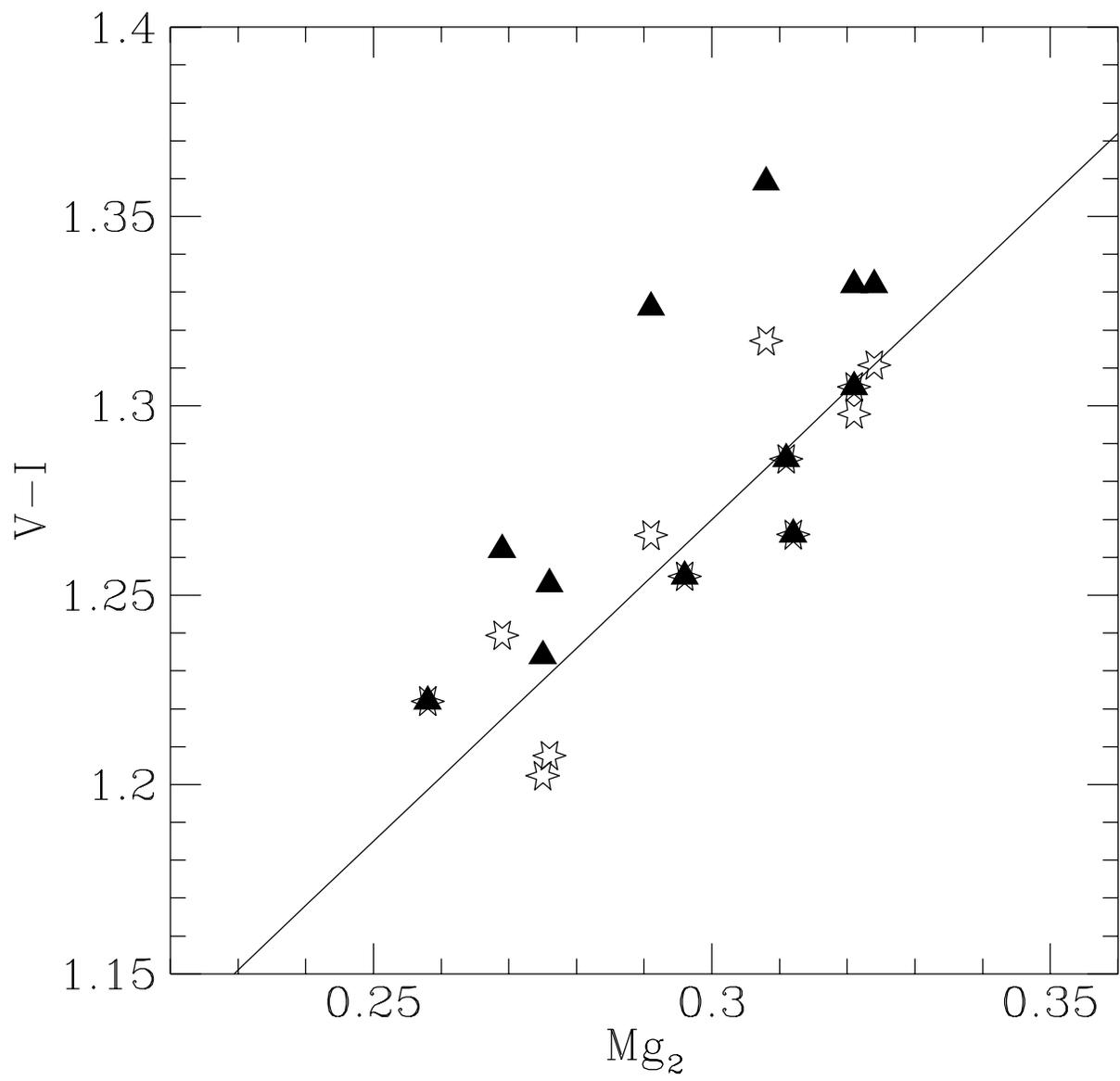
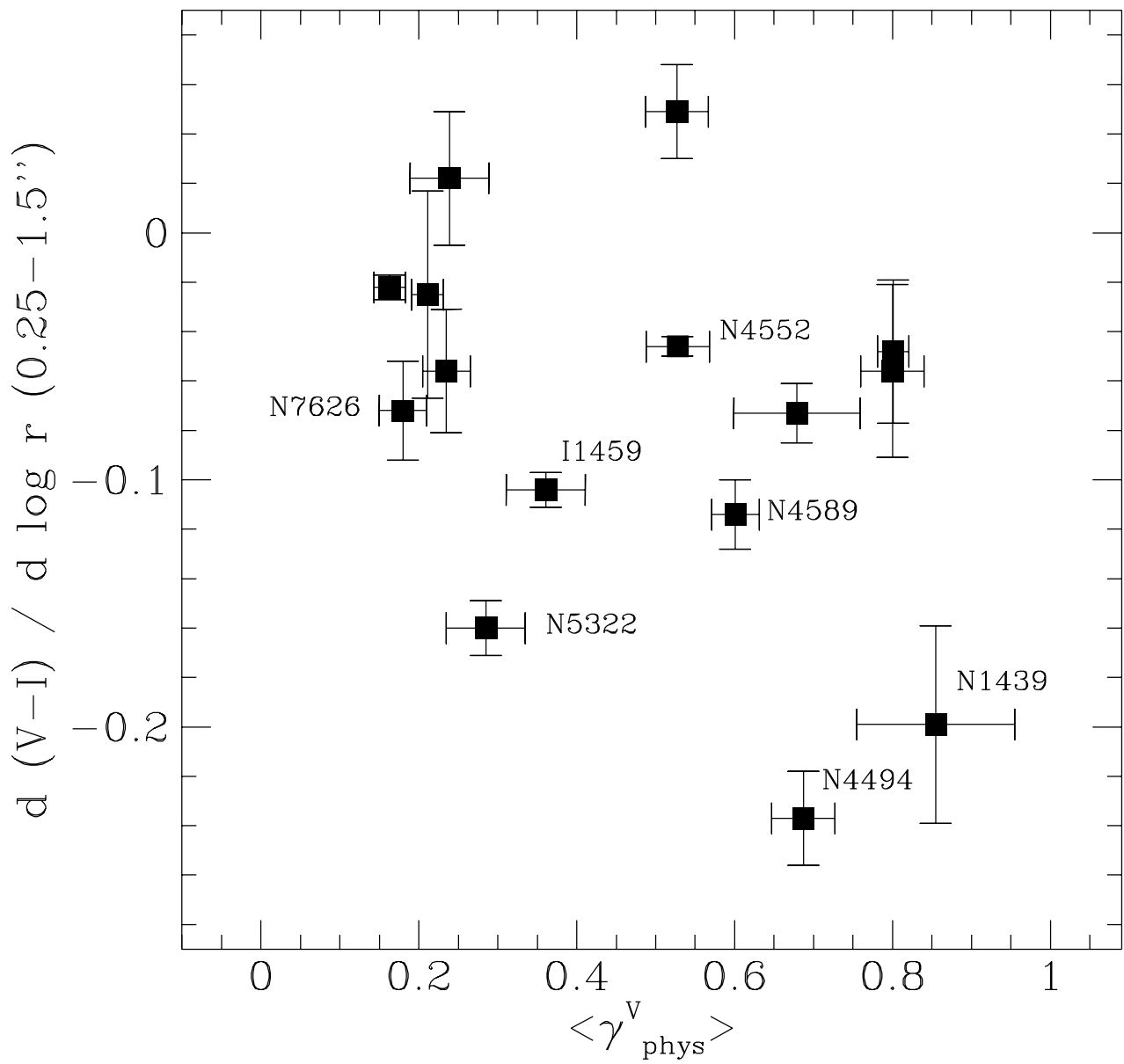
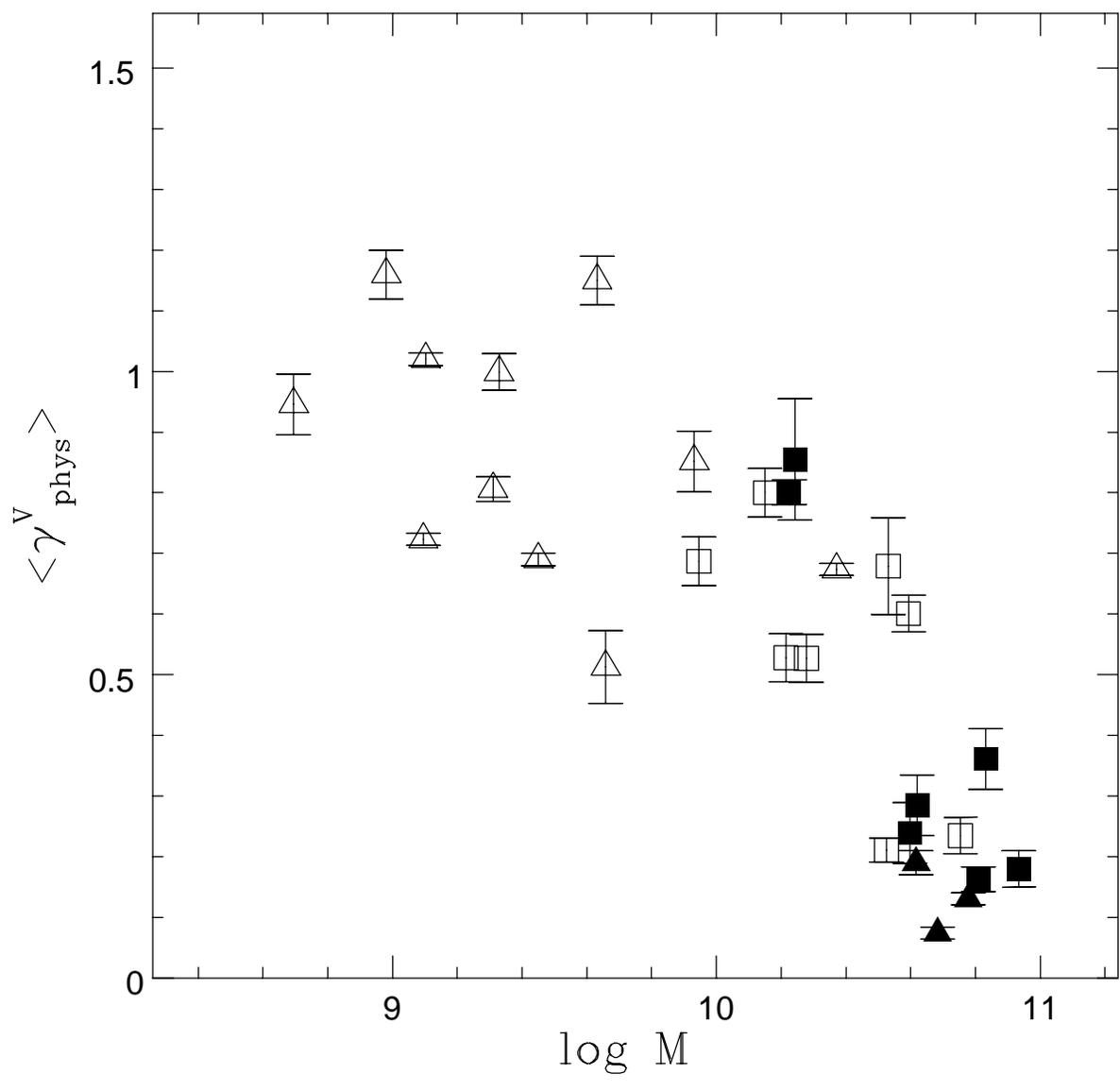
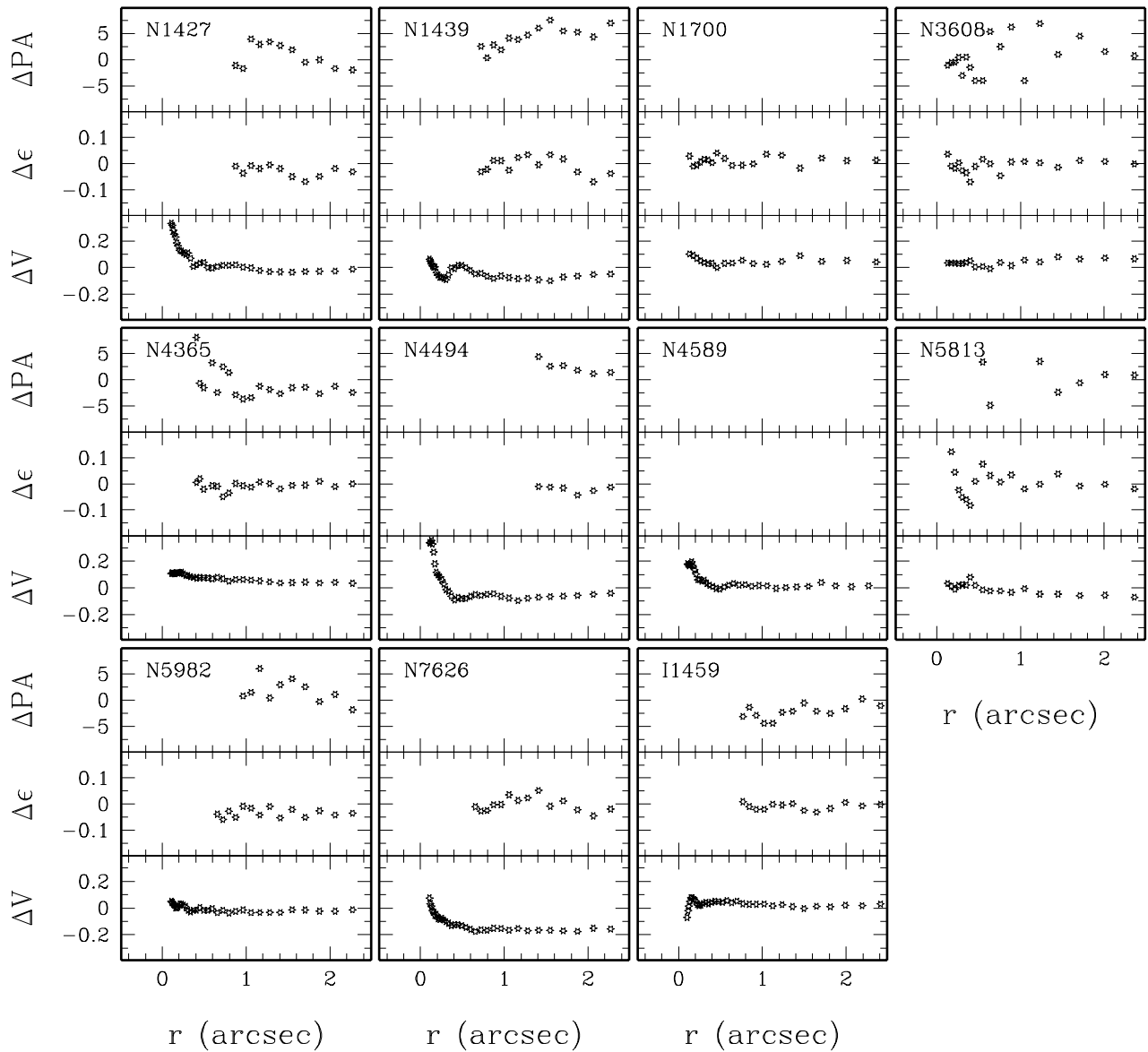
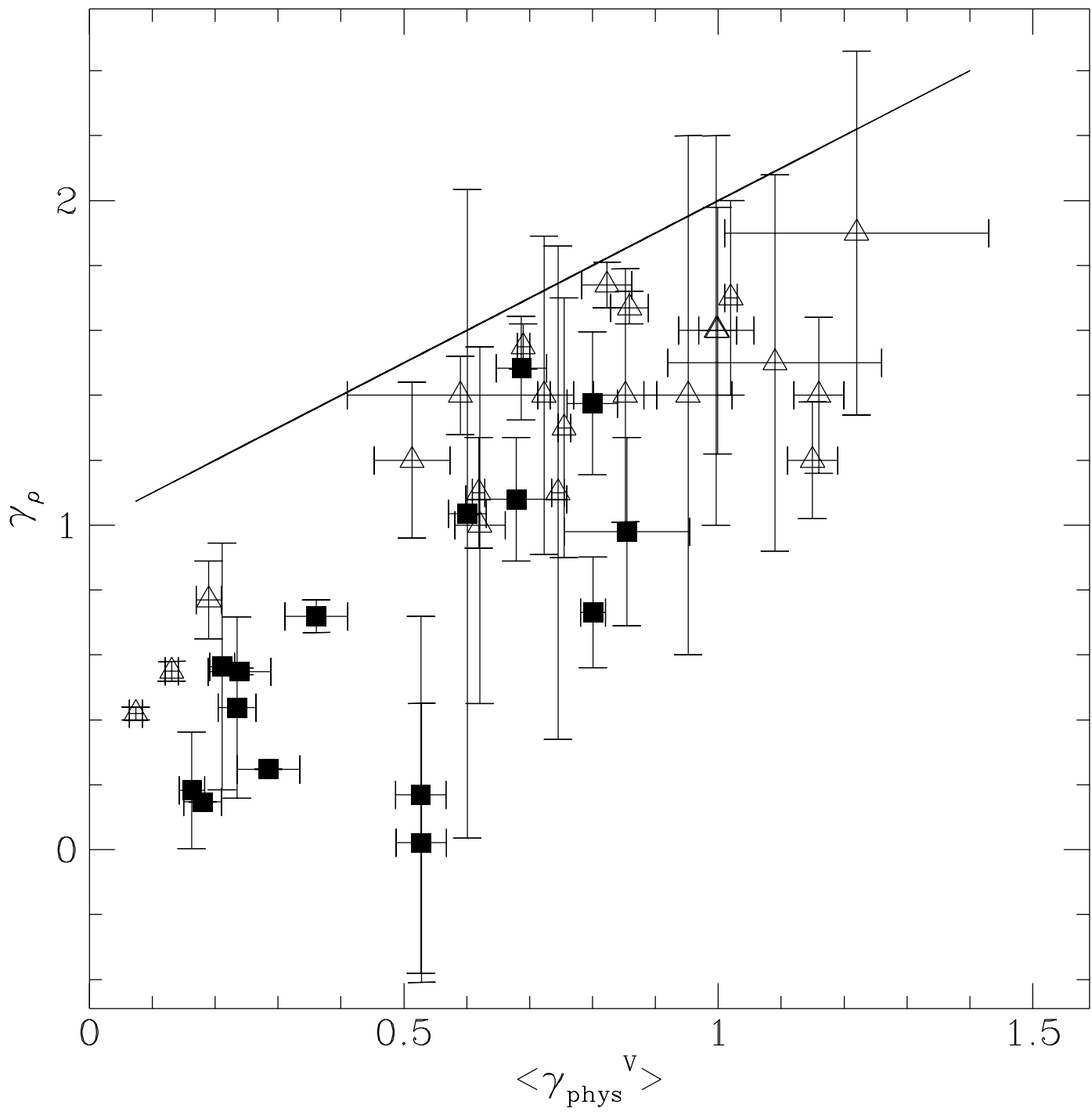
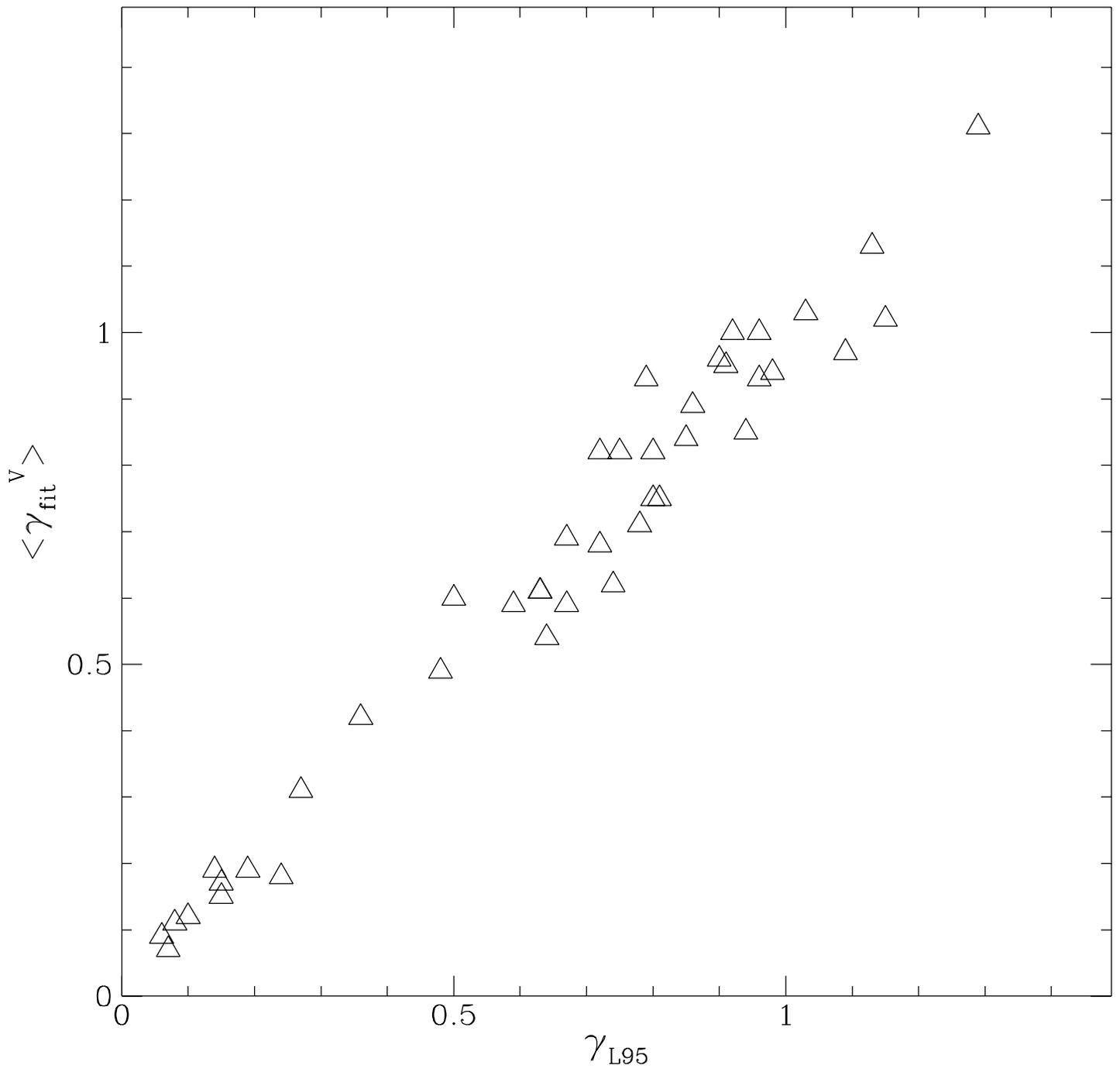
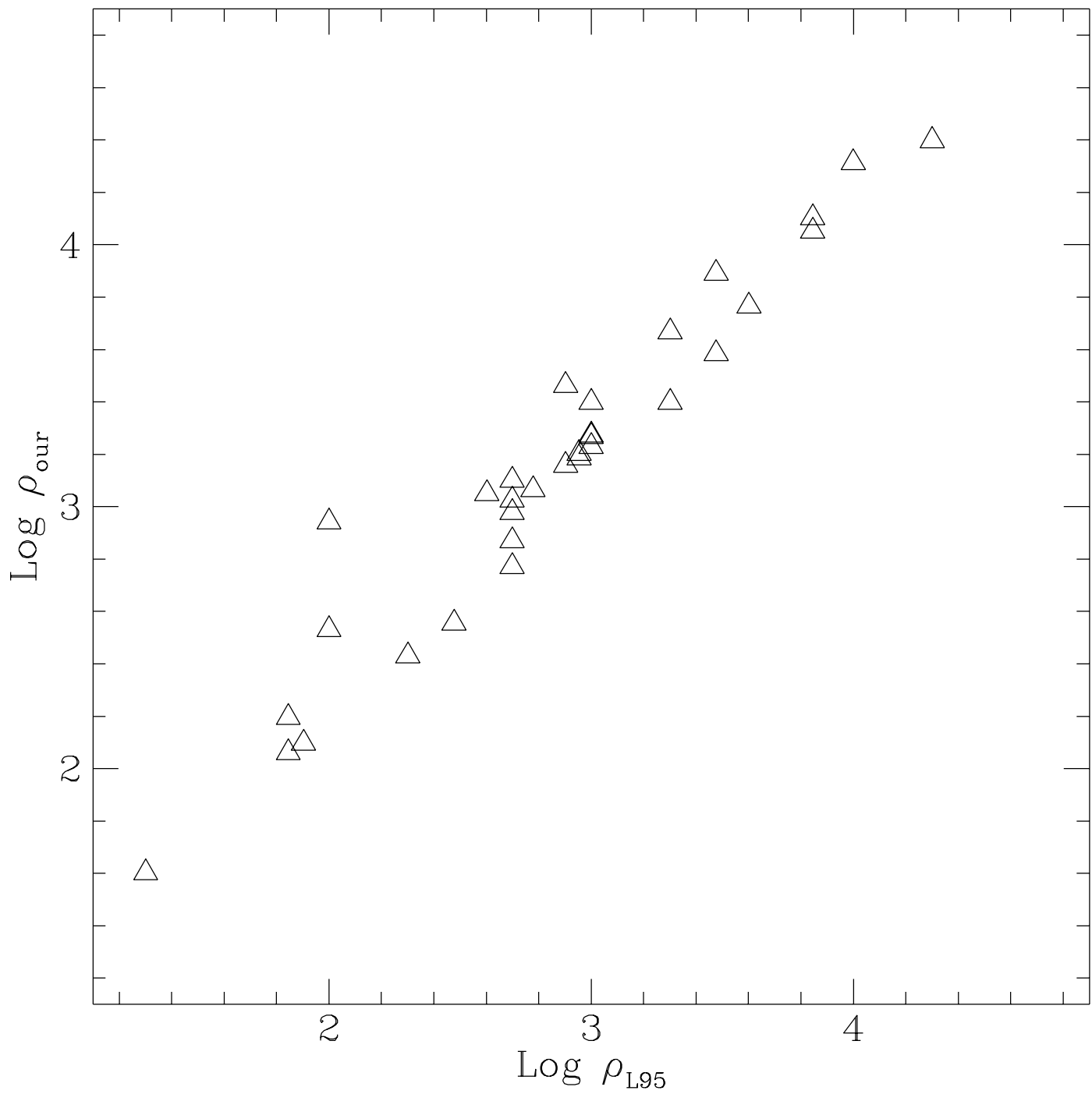
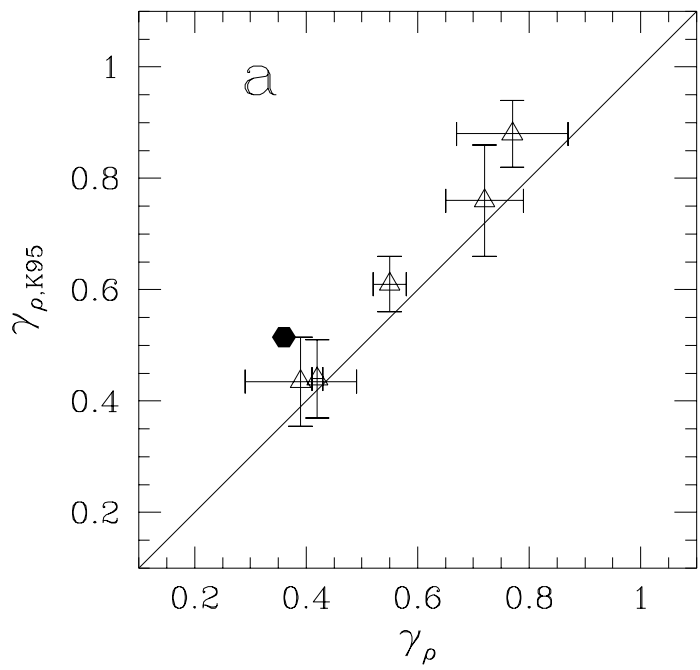
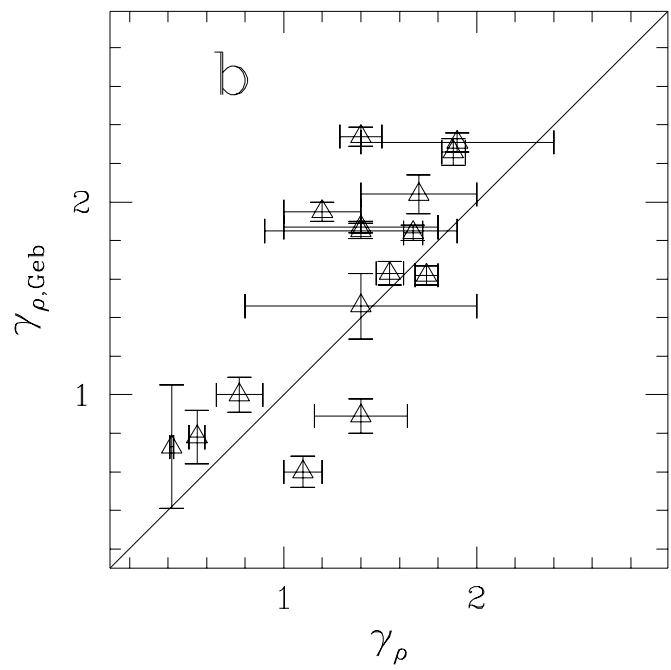
\end{document}